\pdfoutput=1
\RequirePackage{ifpdf}
\ifpdf 
\documentclass[pdftex]{sigma}
\else
\documentclass{sigma}
\fi

\numberwithin{equation}{section}
\newtheorem{Theorem}{Theorem}
\newtheorem{Proposition}[Theorem]{Proposition}

\begin{document}

\renewcommand{\atop}[2]{\genfrac{}{}{0pt}{}{#1}{#2}}
\def\su{\mathfrak{su}}


\newcommand{\arXivNumber}{1507.01821}

\renewcommand{\PaperNumber}{003}

\FirstPageHeading

\ShortArticleName{Doubling (Dual) Hahn Polynomials: Classif\/ication and Applications}

\ArticleName{Doubling (Dual) Hahn Polynomials:\\ Classif\/ication and Applications}

\Author{Roy {OSTE} and Joris {VAN DER JEUGT}}

\AuthorNameForHeading{R.~Oste and J.~Van der Jeugt}

\Address{Department of Applied Mathematics, Computer Science and Statistics, Ghent University,\\ Krijgs\-laan 281-S9, B-9000 Gent, Belgium}
\Email{\href{mailto:Roy.Oste@UGent.be}{Roy.Oste@UGent.be}, \href{mailto:Joris.VanderJeugt@UGent.be}{Joris.VanderJeugt@UGent.be}}

\ArticleDates{Received July 13, 2015, in f\/inal form January 04, 2016; Published online January 07, 2016}

\Abstract{We classify all pairs of recurrence relations in which two Hahn or dual Hahn polynomials with dif\/ferent parameters appear.
Such couples are referred to as (dual) Hahn doubles.
The idea and interest comes from an example appearing in a f\/inite oscillator model~[Jafarov E.I., Stoilova N.I., Van~der Jeugt J., \textit{J.~Phys.~A: Math. Theor.} \textbf{44} (2011), 265203, 15~pages, arXiv:1101.5310].
Our classif\/ication shows there exist three dual Hahn doubles and four Hahn doubles.
The same technique is then applied to Racah polynomials, yielding also four doubles.
Each dual Hahn (Hahn, Racah) double gives rise to an explicit new set of symmetric orthogonal polynomials related to the Christof\/fel and Geronimus transformations.
For each case, we also have an interesting class of two-diagonal matrices with closed form expressions for the eigenvalues.
This extends the class of Sylvester--Kac matrices by remarkable new test matrices.
We examine also the algebraic relations underlying the dual Hahn doubles, and discuss their usefulness for the construction of new f\/inite oscillator models.}


\Classification{33C45; 33C80; 81R05; 81Q65}

\section{Introduction} \label{sec:Introduction}

The tridiagonal $(N+1)\times(N+1)$ matrix of the following form
\begin{gather}
C_{N+1} = \left(
\begin{matrix}
0 & 1 & & & &\\
N & 0 & 2 & & &\\
 & N-1 & 0 & 3 & & \\
 & & \ddots & \ddots & \ddots & \\
 & & & 2 & 0 & N \\
 & & & & 1 & 0
\end{matrix}
\right)
\label{Kac}
\end{gather}
appears in the literature under several names: the Sylvester--Kac matrix, the Kac matrix, the Clement matrix, $\ldots$.
It was already considered by Sylvester~\cite{Sylvester}, used by M.~Kac in some of his seminal work~\cite{Kac}, by Clement as a test matrix for eigenvalue computations~\cite{Clement}, and continues to attract attention~\cite{Bevilacqua, Boros, Taussky}.
The main property of the matrix $C_{N+1}$ is that its eigenvalues are given explicitly by
\begin{gather}
-N, -N+2, -N+4, \dots, N-2, N.
\label{KacEig}
\end{gather}
Because of this simple property, $C_{N+1}$ is a standard test matrix for numerical eigenvalue computations, and part of some standard test matrix toolboxes (e.g.,~\cite{Higham}).

One of the outcomes of the current paper implies that $C_{N+1}$ has appealing two-parameter extensions.
For odd dimensions, let us consider the following tridiagonal matrix
\begin{gather}
C_{2N+1}(\gamma,\delta) = \left(
\begin{matrix}
0 & 2\gamma+2 & & & & & &\\
2N & 0 & 2 & & & & &\\
& 2\delta+2N & 0 & 2\gamma+4 & & & &\\
& & 2N-2 & 0 & 4 & & &\\
& & &\ddots & \ddots & \ddots & &\\
& & & & 2\delta+4 & 0 & 2\gamma+2N & \\
& & & & & 2 & 0 & 2N \\
& & & & & & 2\delta+2 & 0
\end{matrix}
\right).
\label{Kac-odd}
\end{gather}
In the following, we shall sometimes use the term ``two-diagonal''~\cite{BI} for tridiagonal matrices with zero entries on the diagonal
(not to be confused with a bidiagonal matrix, which has also two non-zero diagonals, but for a bidiagonal matrix the non-zero entries are
on the main diagonal and either superdiagonal or the subdiagonal).
So, just as $C_{2N+1}$ the matrix~\eqref{Kac-odd} is two-diagonal, but the superdiagonal of~$C_{2N+1}$,
\begin{gather*}
[1, 2, 3, 4, \ldots, 2N-1, 2N]
\end{gather*}
is replaced by
\begin{gather*}
[2\gamma+2, 2, 2\gamma+4, 4, \ldots, 2\gamma+2N, 2N],
\end{gather*}
and in the subdiagonal of $C_{2N+1}$,
\begin{gather*}
[2N, 2N-1, 2N-2, \ldots, 3, 2, 1]
\end{gather*}
the odd entries are replaced, leading to
\begin{gather*}
[2N, 2\delta+2N, 2N-2, \ldots, 2\delta+4, 2, 2\delta+2].
\end{gather*}
Clearly, for $\gamma=\delta=-\frac12$ the matrix $C_{2N+1}(\gamma,\delta)$ just reduces to $C_{2N+1}$.
One of our results is that $C_{2N+1}(\gamma,\delta)$ has simple eigenvalues for general $\gamma$ and $\delta$,
given by
\begin{gather*}
0, \pm 2\sqrt{1(\gamma+\delta+2)}, \pm 2\sqrt{2(\gamma+\delta+3)}, \pm 2\sqrt{3(\gamma+\delta+4)}, \ldots,
\pm 2\sqrt{N(\gamma+\delta+N+1)}.
\end{gather*}
This spectrum simplif\/ies even further for $\delta=-\gamma-1$; in this case one gets back the eigenva\-lues~\eqref{KacEig}.

For even dimensions, we have a similar result. Let $C_{2N}(\gamma,\delta)$ be the
$(2N)\times(2N)$ tridiagonal matrix with zero diagonal, with superdiagonal
\begin{gather*}
[2\gamma+2, 2, 2\gamma+4, 4, \ldots, 2N-2, 2\gamma+2N]
\end{gather*}
and with subdiagonal
\begin{gather*}
[2\delta+2N, 2N-2, 2\delta+2N-2, \ldots, 4, 2\delta+4, 2, 2\delta+2].
\end{gather*}
Then $C_{2N}(\gamma,\delta)$ has simple eigenvalues for general $\gamma$ and $\delta$,
given by\footnote{In this paper, we shall usually take $\gamma>-1$ and $\delta>-1$, yielding real eigenvalues
	for the matrices under consideration. But the expressions for the eigenvalues remain valid even when these conditions
	are not satisf\/ied.}
\begin{gather}
\pm 2\sqrt{(\gamma+1)(\delta+1)}, \pm 2\sqrt{(\gamma+2)(\delta+2)}, \ldots, \pm 2\sqrt{(\gamma+N)(\delta+N)}.
\label{Kac-even-Eig}
\end{gather}
This spectrum simplif\/ies for $\delta=\gamma$, and obviously for $\gamma=\delta=-\frac12$
one gets back the eigen\-va\-lues~\eqref{KacEig} since in that case $C_{2N}(\gamma,\delta)$ just reduces to~$C_{2N}$.

What is the context here for these new tridiagonal matrices with simple eigenvalue properties?
Well, remember that $C_{N+1}$ also appears as the simplest example of a family of Leonard pairs~\mbox{\cite{Nomura,Terwilliger}}.
In that context, this matrix is related to symmetric Krawtchouk polyno\-mials~\cite{Ismail, Koekoek,Suslov}.
Indeed, let $K_n(x)\equiv K_n(x;\frac12,N)$, where $K_n(x;p,N)$ are the Krawtchouk poly\-no\-mials~\cite{Ismail, Koekoek,Suslov}.
Then their recurrence relation~\cite[equation~(9.11.3)]{Koekoek} yields
\begin{gather}
n K_{n-1}(x) + (N-n) K_{n+1}(x) = (N-2x) K_n(x), \qquad n=0,1,\ldots,N.
\label{Kraw-recur}
\end{gather}
Writing this down for $x=0,1,\ldots,N$, and putting this in matrix form, shows indeed that the eigenvalues of $C_{N+1}$ (or rather,
of its transpose~$C_{N+1}^T$)
are indeed given by~\eqref{KacEig}. Moreover, it shows that the components of the $k$th eigenvector of~$C_{N+1}^T$ are given by~$K_n(k)$.

So we can identify the matrix $C_{N+1}$ with the Jacobi matrix of symmetric Krawtchouk polynomials, one of the families of f\/inite and discrete hypergeometric orthogonal polynomials.
The other matrices $C_{N}(\gamma,\delta)$ appearing in this introduction are not directly related to Jacobi matrices of a simple set of f\/inite orthogonal polynomials.
In this paper, however, we show how two sets of distinct dual Hahn polynomials~\cite{Ismail, Koekoek,Suslov} can be combined in an appropriate way such that the eigenvalues of matrices like~$C_N(\gamma,\delta)$ become apparent, and such that the eigenvector components are given in terms of these two dual Hahn polynomials.
This process of combining two distinct sets is called ``doubling''.
We examine this not only for the case related to the mat\-rix~$C_N(\gamma,\delta)$, but stronger: we classify all possible ways in which two sets of dual Hahn polynomials can be combined in order to yield a two-diagonal ``Jacobi matrix''. It turns out that there are exactly three ways in which dual Hahn polynomials can be ``doubled'' (for a precise formulation, see later).
By the doubling procedure, one automatically gets the eigenvalues (and eigenvectors) of the corresponding two-diagonal matrix in explicit form.

This process of doubling and investigating the corresponding two-diagonal Jacobi matrix can be applied to other classes of orthogonal polynomials (with a~f\/inite and discrete support) as well. In this paper, we turn our attention also to Hahn and to Racah polynomials. The classif\/ication process becomes rather technical, however.
Therefore, we have decided to present the proof of the complete classif\/ication only for dual Hahn polynomials (Section~\ref{sec:dHahn}).
For Hahn polynomials (Section~\ref{sec:Hahn}) we give the f\/inal classif\/ication and corresponding two-diagonal matrices (but omit the proof), and for Racah polynomials we give the f\/inal classif\/ication and some examples of two-diagonal matrices in Appendix~\ref{sec:Racah}.

We should also note that the two-diagonal matrices appearing as a result of the doubling process are symmetric.
So matrices like~\eqref{Kac-odd} do not appear directly but in their symmetrized form.
Of course, as far as eigenvalues are concerned, this makes no dif\/ference (see Section~\ref{sec:testmatrix}).

The doubling process of the polynomials considered here also gives rise to ``new'' sets of orthogonal polynomials.
One could argue whether the term ``new'' is appropriate, since they arise by combining two known sets.
The peculiar property is however that the combined set has a common unique weight function.
Moreover, we shall see that the support set of these doubled polynomials is interesting, see the examples in Section~\ref{sec:orthpoly}.
In this section, we also interpret the doubling process in the framework of Christof\/fel--Geronimus transforms.
It will be clear that from our doubling process, one can deduce for which Christof\/fel parameter the Christof\/fel transform of a Hahn, dual Hahn or Racah polynomial is again a Hahn, dual Hahn or Racah polynomial with shifted parameters.

In Section~\ref{sec:testmatrix} we reconsider the two-diagonal matrices that have appeared in the previous sections.
It should be clear that we get several classes of two-diagonal matrices (with parameters) for which the eigenvalues (and eigenvectors) have an explicit and rather simple form.
This section reviews such matrices as new and potentially interesting examples of eigenvalue test matrices.

In Section~\ref{sec:oscillators} we explore relations with other structures.
Recall that in f\/inite-dimensional representations of the Lie algebra $\su(2)$, with common generators~$J_+$, $J_-$ and $J_0$, the matrix of~\mbox{$J_+ +J_-$} also has a symmetric two-diagonal form.
The new two-diagonal matrices appearing in this paper can be seen as representation matrices of deformations or extensions of~$\su(2)$.
We give the algebraic relations that follow from the ``representation matrices'' obtained here.
The algebras are not studied in detail, but it is clear that they could be of interest on their own.
The general algebras have two parameters, and we indicate how special cases with only one parameter are of importance for the construction of f\/inite oscillator models.

\section{Introductory example} \label{sec:Example}%

We start our analysis by the explanation of a known example taken from~\cite{Stoilova2011}.
For this example, we f\/irst recall the def\/inition of Hahn and dual Hahn polynomials and some of the classical notations and properties.

The Hahn polynomial $Q_n(x;\alpha, \beta, N)$~\cite{Ismail, Koekoek,Suslov} of degree $n$, $n=0,1,\ldots,N$, in the va\-riab\-le~$x$, with parameters
$\alpha>-1$ and $\beta>-1$ (or $\alpha<-N$ and $\beta<-N$)
is def\/ined by~\cite{Ismail, Koekoek,Suslov}
\begin{gather}
Q_n(x;\alpha,\beta,N) = {}_3F_2 \left( \atop{-n,n+\alpha+\beta+1,-x}{\alpha+1,-N} ; 1 \right).
\label{defQ}
\end{gather}
Herein, the function $_3F_2$ is the generalized hypergeometric series~\cite{Bailey,Slater}
\begin{gather}
 {}_3F_2 \left( \atop{a,b,c}{d,e} ; z \right)=\sum_{k=0}^\infty \frac{(a)_k(b)_k(c)_k}{(d)_k(e)_k}\frac{z^k}{k!}.
\label{defF}
\end{gather}
In~(\ref{defQ}), the series is terminating
because of the appearance of the negative integer $-n$ as a numerator parameter.
Note that in~(\ref{defF}) we use the common notation for Pochhammer symbols~\cite{Bailey,Slater}
$(a)_k=a(a+1)\cdots(a+k-1)$ for $k=1,2,\ldots$ and $(a)_0=1$.
Hahn polynomials satisfy a (discrete) orthogonality relation~\cite{Ismail, Koekoek}
\begin{gather}
\sum_{x=0}^N w(x;\alpha, \beta,N) Q_n(x;\alpha, \beta, N) Q_{n'}(x;\alpha,\beta,N) = h_n(\alpha,\beta,N)\, \delta_{n,n'},
\label{orth-Q}
\end{gather}
where
\begin{gather*}
 w(x;\alpha, \beta,N) = \binom{\alpha+x}{x} \binom{N+\beta-x}{N-x}, \qquad x=0,1,\ldots,N, \\
 h_n (\alpha,\beta,N)= \frac{(-1)^n(n+\alpha+\beta+1)_{N+1}(\beta+1)_n n!}{(2n+\alpha+\beta+1)(\alpha+1)_n(-N)_n N!}.
\end{gather*}
We denote the orthonormal Hahn functions as follows
\begin{gather*}
\tilde Q_n(x;\alpha,\beta,N) \equiv \frac{\sqrt{w(x;\alpha,\beta,N)}\, Q_n(x;\alpha,\beta,N)}{\sqrt{h_n(\alpha,\beta,N)}}.
\end{gather*}
The Hahn polynomials satisfy the following recurrence relation~\cite[equation~(9.5.3)]{Koekoek}
\begin{gather}
\label{re}
\Lambda ( x) y_n(x) = A(n) y_{n+1}(x) - \bigl( A(n) + C(n)\bigr) y_n(x) + C(n) y_{n-1}(x)
\end{gather}
with
\begin{gather}
 y_n(x) = Q_n(x;\alpha,\beta,N),\qquad \Lambda(x) = -x, \label{ABD}\\
 A(n) = \frac{(n+\alpha+1)(n+\alpha+\beta+1)(N-n)}{(2n+\alpha+\beta+1)(2n+\alpha+\beta+2)},\qquad
C(n) = \frac{n(n+\alpha+\beta+N+1)(n+\beta)}{(2n+\alpha+\beta)(2n+\alpha+\beta+1)}.\nonumber
\end{gather}

Related to the Hahn polynomials are the dual Hahn polynomials: $R_n(\lambda(x);\gamma, \delta, N)$ of deg\-ree~$n$, $n=0,1,\ldots,N$, in the variable $\lambda(x)=x(x+\gamma+\delta+1)$,
with parameters $\gamma>-1$ and $\delta>-1$ (or $\gamma<-N$ and $\delta<-N$)
which are def\/ined similarly to \eqref{defQ}~\cite{Ismail, Koekoek,Suslov}
\begin{gather}\label{DHahn}
R_n(\lambda(x);\gamma,\delta,N) = {}_3F_2 \left( \atop{-x,x+\gamma+\delta+1,-n}{\gamma+1,-N} ; 1 \right).
\end{gather}
As is well known, the (discrete) orthogonality relation of the dual Hahn polynomials is just the ``dual'' of~\eqref{orth-Q}
\begin{gather}
\sum_{x=0}^N \overline{w}(x;\gamma, \delta,N) R_n(\lambda(x);\gamma, \delta, N) R_{n'}(\lambda(x);\gamma,\delta,N) = \overline{h}_n(\gamma,\delta,N) \delta_{n,n'},
\label{orth-R}
\end{gather}
where
\begin{gather*}
 \overline{w}(x;\gamma, \delta,N) = \frac{(2x+\gamma+\delta+1)(\gamma+1)_x(-N)_x N!}{(-1)^x(x+\gamma+\delta+1)_{N+1}(\delta+1)_x x!},\\
 \overline{h}_n (\gamma,\delta,N)= \left[\binom{\gamma+n}{n} \binom{N+\delta-n}{N-n}\right]^{-1} .
\end{gather*}
Orthonormal dual Hahn functions are def\/ined by
\begin{gather}
\tilde R_n(\lambda(x);\gamma,\delta,N) \equiv \frac{\sqrt{\overline{w}(x;\gamma,\delta,N)} R_n(\lambda(x);\gamma,\delta,N)}{\sqrt{\overline{h}_n(\gamma,\delta,N)}}.
\label{R-tilde}
\end{gather}
Dual Hahn polynomials also satisfy a recurrence relation of the form~\eqref{re},
with~\cite[equation~(9.6.3)]{Koekoek}
\begin{gather}
 y_n(x) = R_n(\lambda ( x);\gamma,\delta,N),\qquad \Lambda(x)=\lambda(x) = x(x+\gamma+\delta+1),\nonumber\\
 A(n) = (n+\gamma+1)(n-N),\qquad C(n) = n(n-\delta-N-1).
\label{AC}
\end{gather}

In~\cite{Stoilova2011}, the following dif\/ference equations involving two sets of Hahn polynomials were derived
(for convenience we use the notation $Q_n(x)\equiv Q_n(x;\alpha,\beta+1,N)$ and $\hat Q_n(x)\equiv Q_n(x;\alpha+1,\beta,N)$):
\begin{gather}
 (N+\beta+1-x) Q_n(x)-(N-x) Q_{n}(x+1) =\frac{(n+\alpha +1)(n+\beta+1)}{\alpha+1} \hat Q_n(x), \label{Q-rec1} \\
 (x+1) \hat Q_n(x)-(\alpha +x+2) \hat Q_{n}(x+1) =-(\alpha+1) Q_n(x+1). \label{Q-rec2}
\end{gather}
Writing out these dif\/ference equations for $x=0,1,\ldots,N$, the resulting set of equations can easily be written in matrix form.
For this matrix form, let us use the normalized version of the polynomials, and construct the following $(2N+2)\times (2N+2)$ matrix $U$ with elements
\begin{gather}
 U_{2x,N-n} = U_{2x,N+n+1} = \frac{(-1)^x}{\sqrt{2}} \tilde Q_n(x;\alpha,\beta+1,N), \label{Ueven}\\
 U_{2x+1,N-n} = -U_{2x+1,N+n+1} = -\frac{(-1)^x}{\sqrt{2}} \tilde Q_n(x;\alpha+1,\beta,N), \label{Uodd}
\end{gather}
where $x,n\in\{0,1,\ldots,N\}$.
By construction, this matrix is orthogonal~\cite{Stoilova2011}: the fact that the columns of $U$ are orthonormal follows from the orthogonality relation of the Hahn polynomials, and from the signs in the matrix $U$. Thus $U^TU=UU^T=I$, the identity matrix.

The normalized dif\/ference equations~\eqref{Q-rec1},~\eqref{Q-rec2} for $x=0,1,\ldots,N$ can then be cast in matrix form.
The coef\/f\/icients in the left hand sides of~\eqref{Q-rec1},~\eqref{Q-rec2} give rise to a~tridiagonal $(2N+2)\times(2N+2)$-matrix of the form
\begin{gather}
\label{MK}
M= \left( \begin{matrix}
 0 & M_0 & 0 & & \\
 M_0 & 0 & M_1 & \ddots & \\
 0 & M_1 & 0 & \ddots & 0 \\
 &\ddots & \ddots & \ddots & M_{2N} \\
 & & 0 & M_{2N} & 0
 \end{matrix} \right),
\end{gather}
 with
 \begin{gather}
 M_{2k}= \sqrt{(k+\alpha+1)(N+\beta+1-k)}, \qquad M_{2k+1}=\sqrt{(k+1)(N-k)}.
 \label{M_k}
 \end{gather}
Suppose $\alpha>-1$, $\beta>-1$ or $\alpha<-N-1$, $\beta<-N-1$ and let $U$ be the orthogonal matrix determined in~\eqref{Ueven},~\eqref{Uodd}.
Then~\cite{Stoilova2011} the columns of $U$ are the eigenvectors of $M$, i.e.,
\begin{gather}
M U = U D,
\label{MUUD}
\end{gather}
where $D$ is a diagonal matrix containing the eigenvalues of~$M$
\begin{gather}
 D= \operatorname{diag} (-\epsilon_N,\ldots,-\epsilon_1,-\epsilon_0,\epsilon_0,\epsilon_1,\ldots,\epsilon_{N}), \nonumber\\
 \epsilon_{k}=\sqrt{(\alpha+k+1)(\beta+k+1)},\qquad k=0,1,\ldots,N. \label{epsilon}
\end{gather}

Note that the eigenvalues of the matrix $M$ are (up to a factor~2) the same as those of the matrix
$C_{2N+2}(\alpha,\beta)$, the two-parameter extension of the Sylvester--Kac matrix. As we will further discuss in Section~\ref{sec:testmatrix}, the above result proves that the eigenvalues of $C_{2N+2}(\alpha,\beta)$ are indeed given by~\eqref{Kac-even-Eig}.
Even more: the orthonormal eigenvectors of~$M$ are just the columns of~$U$.

Another way of looking at~\eqref{MUUD} is in terms of the dual Hahn polynomials. Interchanging~$x$ and~$n$ in the expressions~\eqref{Ueven},~\eqref{Uodd}, we have
\begin{gather}
 U_{2n,N-x} = U_{2n,N+x+1} = \frac{(-1)^n}{\sqrt{2}} \tilde R_n(\lambda(x);\alpha,\beta+1,N), \label{UevenR}\\
 U_{2n+1,N-x} = -U_{2n+1,N+x+1} = -\frac{(-1)^n}{\sqrt{2}} \tilde R_n(\lambda(x);\alpha+1,\beta,N), \label{UoddR}
\end{gather}
where $x,n\in\{0,1,\ldots,N\}$. In this way, each row of the matrix $U$ consists of a dual Hahn polynomial of a certain degree, having dif\/ferent parameter values for even and odd rows.
Now, the relation~\eqref{MUUD} can be interpreted as a three-term recurrence relation with $M$ being the Jacobi matrix.
Two sets of (dual) Hahn polynomials (with dif\/ferent parameters) are thus combined into a new set of polynomials such that the Jacobi matrix for this new set has a simple two-diagonal form, with simple eigenvalues.
The pair of dif\/ference equations~\eqref{Q-rec1},~\eqref{Q-rec2} involving two sets of Hahn polynomials then corresponds to the following relations involving the dual Hahn polynomials $R_n(x)\equiv R_{n}(\lambda(x);\gamma,\delta+1,N)$ and $\hat R_n(x)\equiv R_{n}(\lambda(x);\gamma+1,\delta,N)$:
\begin{gather}\label{R-rec1}
	 (N+\delta+1-n)R_{n}(x)-(N-n) R_{n+1}(x) =\frac{(x+\gamma+1)(x+\delta+1)}{(\gamma+1)} \hat R_n(x),\\
	 (n+1) \hat R_n(x)- (n+\gamma+2) \hat R_{n+1}(x) = -(\gamma+1) R_{n+1}(x).\label{R-rec2}
\end{gather}
This is in fact a special case of the so-called Christof\/fel transform of a dual Hahn polynomial with its transformation parameter chosen specif\/ically so that the result is again a dual Hahn polynomial (with dif\/ferent parameters). We will further elaborate on this in Section~\ref{sec:orthpoly}.

This introductory example, taken from~\cite{Stoilova2011}, opens the following question:
in how many ways can two sets of (dual) Hahn polynomials be combined such that the Jacobi matrix is two-diagonal?
This will be answered in the following section.

\section{Doubling dual Hahn polynomials: classif\/ication} \label{sec:dHahn}

The essential relation in the previous example is the existence of a pair of ``recurrence relations''~\eqref{R-rec1},~\eqref{R-rec2} intertwining two types of dual Hahn polynomials (or equivalently a couple of dif\/ference equations~\eqref{Q-rec1},~\eqref{Q-rec2} for two types of their duals, the Hahn polynomials).
Let us therefore examine the existence of such relations in general.
Say we have two types of dual Hahn polynomials with dif\/ferent parameter values for $\gamma$ and $\delta$ (and possibly $N$) denoted by $R_n(\lambda ( x);\gamma,\delta,N)$ and $R_{n}(\lambda (\hat x);\hat{\gamma},\hat{\delta},\hat{N})$, that are related in the following manner
\begin{gather}
\label{rel1}
a(n) R_n(\lambda ( x);\gamma,\delta,N) + b(n) R_{n+1}(\lambda ( x);\gamma,\delta,N) ={\hat{d}}(x) R_{n}(\lambda (\hat x);\hat{\gamma},\hat{\delta},\hat{N}) \\
\hat{a}(n) R_{n}(\lambda (\hat x);\hat{\gamma},\hat{\delta},\hat{N}) + \hat{b}(n) R_{n+1}(\lambda (\hat x);\hat{\gamma},\hat{\delta},\hat{N}) = d(x) R_{n+1}(\lambda ( x);\gamma,\delta,N) .
\label{rel2}
\end{gather}
If we want these relations to correspond to a matrix identity like~\eqref{MUUD}, then it is indeed necessary that the (unknown) functions~$a(n)$, $\hat a(n)$, $b(n)$ and~$\hat b(n)$ are functions of $n$ and not of~$x$, and that~$d(x)$ and~$\hat d(x)$ are functions of~$x$ and not of~$n$.
Of course, the parameters~$\gamma$, $\delta$, $N$, $\hat\gamma$, $\hat\delta$, $\hat N$ can appear in these functions.

In order to lift this technique also to other polynomials than just the dual Hahn polynomials,
say we have the following relations between two sets of orthogonal polynomials of the same class, denoted by~$y_{n}$ and~$\hat{y}_{n}$, but with dif\/ferent parameter values
\begin{gather}
\label{yyh}
a(n) y_{n} + b(n) y_{n+1} ={\hat{d}}(x) \hat{y}_{n} ,\\
\hat{a}(n) \hat{y}_{n} + \hat{b}(n) \hat{y}_{n+1} = d(x) y_{n+1} ,
\label{hhy}
\end{gather}
where $a$, ${\hat{a}}$, $b$, ${\hat{b}}$ are independent of $x$ and $d$, ${\hat{d}}$ are independent of~$n$. Although~\eqref{yyh},~\eqref{hhy} are not actual recurrence relations since they involve both~$y_{n}$ and~$\hat{y}_{n}$, we will refer to a couple of such relations intertwining two types of orthogonal polynomials as ``a~pair of recurrence relations''.

When substituting~\eqref{hhy} in~\eqref{yyh}, we arrive at the following recurrence relation for~$\hat y_{n}$
\begin{gather}
\label{de1}
a(n) \bigl\lbrack \hat{a}(n-1) \hat{y}_{n-1} + \hat{b}(n-1) \hat{y}_{n}\bigr\rbrack + b(n) \bigl\lbrack \hat{a}(n) \hat{y}_{n} + \hat{b}(n) \hat{y}_{n+1}\bigr\rbrack =d(x) {\hat{d}}(x) \hat{y}_{n}.
\end{gather}
In the same manner, $\hat{y}_{n}$ can be eliminated to f\/ind a recurrence relation for~$y_{n}$
\begin{gather}
\label{deh1}
\hat{a}(n-1) \bigl\lbrack a(n-1) y_{n-1} + b(n-1) y_{n} \bigr\rbrack + \hat{b}(n-1) \bigl\lbrack a(n) y_{n} + b(n) y_{n+1}\bigr\rbrack = {\hat{d}}(x) d(x) y_{n} .
\end{gather}

Of course, the orthogonal polynomials $y_{n}$ already satisfy a three-term recurrence relation of the form~\eqref{re}.
A comparison of the coef\/f\/icients of $y_{n+1}$, $ y_{n}$, $y_{n-1}$ in \eqref{de1},~\eqref{deh1} with the known coef\/f\/icients given in~\eqref{AC} leads to the following set of requirements for $a$, ${\hat{a}}$, $b$, ${\hat{b}}$, $d$, ${\hat{d}}$
\begin{gather}
\label{Dh}
a(n) \hat{a}(n-1) ={\hat{C}}(n), \\
\label{D}
a(n-1) \hat{a}(n-1) =C(n), \\
\label{BDh}
 a(n) \hat{b}(n-1) + \hat{a}(n) b(n) - d(x)\,{\hat{d}}(x) = - \bigl\lbrack \hat{\Lambda} ( x)+ {\hat{A}}(n) + {\hat{C}}(n)\bigr\rbrack, \\
\label{BD}
 a(n) \hat{b}(n-1) + \hat{a}(n-1) b(n-1) - {\hat{d}}(x)d(x) = - \bigl\lbrack \Lambda ( x)+ A(n) + C(n)\bigr\rbrack, \\
\label{Bh}
b(n) \hat{b}(n) = {\hat{A}}(n), \\
\label{B}
 b(n) \hat{b}(n-1) = A(n).
\end{gather}
After a slight rearrangement of terms in the requirements~\eqref{BDh} and~\eqref{BD}, we arrive at two new equations where the left hand side is independent of~$x$ while the right hand side is independent of~$n$, namely,
\begin{gather}
 \label{BDh2}
 a(n) \hat{b}(n-1) + \hat{a}(n) b(n) + {\hat{A}}(n) + {\hat{C}}(n) = d(x){\hat{d}}(x)- \hat{\Lambda} ( x), \\
 \label{BD2}
 a(n) \hat{b}(n-1) + \hat{a}(n-1) b(n-1) + A(n) + C(n) = {\hat{d}}(x) d(x)- \Lambda ( x).
\end{gather}
Hence, the two sides must be independent of both $n$ and $x$. By means of \eqref{Dh}--\eqref{B} we can eliminate $A$, $\hat{A}$, $C$, $\hat{C}$ to f\/ind
 \begin{gather*}
 a(n) \bigl\lbrack \hat{a}(n-1) + \hat{b}(n-1) \bigr\rbrack + b(n)\bigl\lbrack \hat{a}(n) +\hat{b}(n) \bigr\rbrack = d(x) {\hat{d}}(x)- \hat{\Lambda} ( x), \\
 \hat{a}(n-1) \bigl\lbrack a(n-1) + b(n-1) \bigr\rbrack +\hat{b}(n-1) \bigl\lbrack a(n) + b(n)\bigr\rbrack = {\hat{d}}(x) d(x)- \Lambda ( x).
 \end{gather*}
Moreover, subtracting one from the other yields
\begin{gather}
 \Lambda ( x) - \hat{\Lambda} ( x)
 = \hat{a}(n-1) \bigl\lbrack a(n)-a(n-1) - b(n-1) \bigr\rbrack
 + b(n) \bigl\lbrack \hat{a}(n) + \hat{b}(n) - \hat{b}(n-1) \bigr\rbrack. \!\!\!\label{LL}
\end{gather}

Now, for a given class of orthogonal polynomials with recurrence relation of the form~\eqref{re}, we determine all possible functions $a$, $\hat{a}$, $b$, $\hat{b}$, $d$, ${\hat{d}}$ satisfying the list of requirements \eqref{Dh}--\eqref{B}. Hereto, we proceed as follows
\begin{itemize}\itemsep=0pt
\item
From \eqref{Dh} and \eqref{D} we observe that, up to a multiplicative factor, $C(n)$ is split into two functions, $a(n-1)$ and $\hat{a}(n-1)$. When $a(n-1)$ is shifted by~1 in~$n$ and multiplied again by $\hat{a}(n-1)$ we must arrive at~$\hat{C}(n)$. Hence, $C$ and $\hat{C}$ consist of an identical part, and a~part which dif\/fers by a shift of~1 in~$n$. This observation gives a f\/irst list of possibilities for~$a$ and~$\hat{a}$.
\item
Similarly we f\/ind a list for $b$ and $\hat{b}$ by means of \eqref{Bh} and \eqref{B}.
\item
These possibilities are then to be compared with requirements~\eqref{BDh} and~\eqref{BD}.
From \eqref{BDh2},~\eqref{BD2} and~\eqref{LL} we get an expression for the product $d(x) {\hat{d}(x)}$.
Finally, the set of remaining choices for $a$, ${\hat{a}}$, $b$, ${\hat{b}}$ are to be plugged in~\eqref{hhy} and~\eqref{yyh} in order to get~$d$,~${\hat{d}}$ and to verify if these relations indeed hold.
\end{itemize}

The actual performance of the procedure just described is still quite long and tedious, when carried out for a f\/ixed class of polynomials.
In what follows we achieve this for the dual Hahn polynomials, which have the easiest recurrence relation, and it takes about three pages to present this.
The reader who wishes to skip the details can advance to Theorem~\ref{theo2}.

For dual Hahn polynomials, the data is given by~\eqref{AC}
\begin{gather*}
y_n = R_n(\lambda ( x);\gamma,\delta,N),\qquad \hat{y}_n = R_{n}\big(\lambda (\hat x);\hat{\gamma},\hat{\delta},\hat{N}\big), \qquad \Lambda(x) =\lambda ( x)= x(x+\gamma+\delta+1),
\\
A(n) = (n+\gamma+1)(n-N),\qquad
C(n) = n(n-\delta-N-1),
\end{gather*}
and with similar expressions for $\hat\Lambda(x)$, $\hat A(n)$ and $\hat C(n)$ (with $x$, $\gamma$, $\delta$, $N$ replaced by $\hat x$, $\hat\gamma$, $\hat\delta$, $\hat N$).
From~\eqref{LL}, the following expression must be independent of~$x$
\begin{gather*}
\Lambda ( x) - \hat{\Lambda} ( x) = x(x+\gamma+\delta+1) - \hat{x}\big(\hat{x}+\hat{\gamma}+\hat{\delta}+1\big).
\end{gather*}
In order for the term in $x^2$ to disappear, we must have $\hat{x} = x + \xi$ which gives
\begin{gather*}
x(x+\gamma+\delta+1) - ( x + \xi)( x + \xi+\hat{\gamma}+\hat{\delta}+1) = (\gamma+\delta-\hat{\gamma}-\hat{\delta} -2 \xi ) x - \xi( \xi+\hat{\gamma}+\hat{\delta}+1)
\end{gather*}
and as we require the coef\/f\/icient of $x$ to be zero we f\/ind the following condition for~$\xi$
\begin{gather}
\label{ab}
\gamma+\delta-(\hat{\gamma}+\hat{\delta}) = 2 \xi .
\end{gather}

From \eqref{D} we see that we have four distinct possible combinations for~$a(n-1)$ and~$\hat{a}(n-1)$
\begin{alignat*}{3}
& a(n-1) = 1 c_a, \qquad && \hat{a}(n-1) = n(n-\delta-N-1) c_a^{-1}, & \tag{a1}\label{a1}\\
& a(n-1) = n c_a, \qquad && \hat{a}(n-1) = (n-\delta-N-1) c_a^{-1}, & \tag{a2}\label{a2}\\
& a(n-1) = (n-\delta-N-1) c_a, \qquad && \hat{a}(n-1) = n c_a^{-1}, & \tag{a3}\label{a3}\\
& a(n-1) = n(n-\delta-N-1) c_a, \qquad && \hat{a}(n-1) = 1 c_a^{-1}, &\tag{a4}\label{a4}
\end{alignat*}
with $c_a$ a factor.
Combining this with \eqref{Dh} we must have
\begin{gather*}
a(n) \hat{a}(n-1) ={\hat{C}}(n) = n\big(n-\hat{\delta}-\hat{N}-1\big).
\end{gather*}
This immediately implies that $c_a$ is independent of $n$, and \eqref{a1}--\eqref{a4} yield the following possibilities
\begin{align*}
& n(n-\delta-N-1) = n(n-\hat{\delta}-\hat{N}-1) \ && \implies \ \delta+N =\hat{\delta}+\hat{N}, & \tag{a1$'$}\label{a1'}\\
& (n+1) (n-\delta-N-1) = n(n-\hat{\delta}-\hat{N}-1)\!\!\! && \implies\ \delta+N+1 = 0 \, \land\, \hat{\delta}+\hat{N}+2=0,\!\!\!\!\!\!\!\! & \tag{a2$'$}\label{a2'}\\
& (n-\delta-N) n = n(n-\hat{\delta}-\hat{N}-1) \ && \implies \ \delta+N =\hat{\delta}+\hat{N}+1, & \tag{a3$'$}\label{a3'}\\
& (n+1)(n-\delta-N)= n(n-\hat{\delta}-\hat{N}-1) \ && \implies \ \delta+N = 0\, \land \, \hat{\delta}+\hat{N}+2 =0. & \tag{a4$'$}\label{a4'}
\end{align*}
Because of the restriction on $\delta$ the option \eqref{a4'} is ineligible, leaving \eqref{a1'}--\eqref{a3'} as only viable options.

In a similar way, from \eqref{B} we see that we have four possible combinations for~$b(n)$ and~$\hat{b}(n)$,
\begin{alignat*}{3}
& b(n)= 1 c_b,\qquad & & \hat{b}(n-1) = (n+\gamma+1)(n-N) c_b^{-1}, & \tag{b1}\label{b1}\\
& b(n)= (n+\gamma+1) c_b,\qquad & & \hat{b}(n-1) = (n-N) c_b^{-1}, & \tag{b2}\label{b2}\\
& b(n)= (n-N)c_b,\qquad & & \hat{b}(n-1) = (n+\gamma+1) c_b^{-1}, &\tag{b3}\label{b3} \\
& b(n)= (n+\gamma+1)(n-N) c_b,\qquad & & \hat{b}(n-1) = 1 c_b^{-1}. &\tag{b4}\label{b4}
\end{alignat*}
Combining this with \eqref{Bh} we must have
\begin{gather*}
b(n) \hat{b}(n) ={\hat{A}}(n) = (n+\hat{\gamma}+1)(n-\hat{N})
\end{gather*}
This implies that $c_b$ is independent of $n$ and moreover for \eqref{b1}--\eqref{b4} yields
\begin{align*}
& (n+\gamma+2)(n-N+1) = (n+\hat{\gamma}+1)(n-\hat{N}) && \implies \ \gamma+1 =\hat{\gamma} \, \land \, N-1 =\hat{N}, & \tag{b1$'$}\label{b1'}\\
& (n+\gamma+1)(n-N+1) = (n+\hat{\gamma}+1)(n-\hat{N}) && \implies \ \gamma =\hat{\gamma} \, \land \, N-1 =\hat{N}, & \tag{b2$'$}\label{b2'}\\
& (n+\gamma+2)(n-N) = (n+\hat{\gamma}+1)(n-\hat{N}) && \implies \ \gamma+1 =\hat{\gamma} \, \land \, N =\hat{N}, & \tag{b3$'$}\label{b3'}\\
& (n+\gamma+1)(n-N)= (n+\hat{\gamma}+1)(n-\hat{N}) && \implies \ \gamma =\hat{\gamma} \, \land \, N =\hat{N}. & \tag{b4$'$}\label{b4'}
\end{align*}
We thus have four viable options for $b$, $\hat{b}$ and three for $a$, $\hat{a}$, giving a total of 12 possible combinations, which we will systematically consider and treat.

{\em Case \eqref{b1}}.
Plugging \eqref{b1} in \eqref{BD2}, we get
\begin{gather*}
a(n) (n+\gamma+1)(n-N)\, c_b^{-1} + \hat{a}(n-1) c_b + (n+\gamma+1)(n-N)+
 n(n-\delta-N-1)\\
 \qquad{} = {\hat{d}}(x)d(x)- \Lambda ( x).
\end{gather*}
As the right hand side is independent of~$n$, so must be the left hand side. This eliminates op\-tions~\eqref{a2} and \eqref{a3} for $a$, $\hat{a}$ as that would result in a third order term in~$n$ which cannot vanish. On the other hand, \eqref{a1} yields
\begin{gather*}
 (n+\gamma+1)(n-N) \frac{c_a}{ c_b} + n(n-\delta-N-1)\frac{c_b}{ c_a} + (n+\gamma+1)(n-N)+
n(n-\delta-N-1) \\
\qquad{} = {\hat{d}}(x)d(x)- \Lambda ( x).
\end{gather*}
This must be independent of $n$, so the coef\/f\/icient of~$n^2$ in the left hand side must vanish, hence ${c_a}/{ c_b} + {c_b}/{ c_a} +2=0$ or thus ${c_a}/{ c_b}=-1$.
For this value of ${c_a}/{ c_b}$ the left hand side equals zero and is indeed independent of~$n$.
Note that this leaves one degree of freedom as only the ratio ${c_a}/{ c_b}$ is f\/ixed. This is just a global scalar factor for \eqref{yyh} and \eqref{hhy}, also present in~\eqref{re}. Henceforth, for convenience, we set $c_a =1$ and $c_b=-1$.

The combined options \eqref{b1} and~\eqref{a1} thus give a valid set of equations of the form~\eqref{yyh} and~\eqref{hhy}, and they correspond to the parameter values
\begin{gather*}
\hat{\gamma} = \gamma+1,\qquad \hat{\delta}= \delta+1,\qquad \hat{N} = N-1.
\end{gather*}
Moreover, by means of \eqref{ab} we f\/ind $\xi=-1$ and so $\hat{x} = x-1$.
Finally, plugging these $a$, ${\hat{a}}$, $b$,~${\hat{b}}$ in~\eqref{yyh} and~\eqref{hhy}, and putting $n=0$ we f\/ind
\begin{gather*}
R_{0}(\lambda(x);\gamma,\delta,N)- R_{1}(\lambda(x);\gamma,\delta,N) = \frac{x(x+\gamma+\delta+1)}{N(\gamma+1)} = \hat{d}(x)
\end{gather*}	
and similarly $d(x)= N(\gamma+1)$.
Hence, for $R_n(x)\equiv R_n(\lambda(x);\gamma,\delta,N)$ and $\hat R_n(x)\equiv R_n(\lambda(x-1)$; $\gamma+1,\delta+1,N-1)$ we have the relations
\begin{gather*}
 R_{n}(x)- R_{n+1}(x) = \frac{x(x+\gamma+\delta+1)}{N(\gamma+1)} \hat R_{n}(x), \\
 -(n+1)(N-n+\delta) \hat R_{n}(x)+(N-n-1)(n+\gamma+2) \hat R_{n+1}(x) = N(\gamma+1)R_{n+1}(x).
\end{gather*}
Interchanging $x$ and $n$, these recurrence relations for dual Hahn polynomials are precisely the known actions of the forward and backward shift operator for Hahn polynomials~\cite[equations~(9.5.6) and~(9.5.8)]{Koekoek}.

 {\em Case \eqref{b2}}.
Next, we consider the option \eqref{b2} for $b$, $\hat{b}$. Plugging \eqref{b2} in \eqref{BD2}, we get
\begin{gather*}
a(n) (n-N) c_b^{-1} + \hat{a}(n-1) (n+\gamma) c_b + (n+\gamma+1)(n-N)+
n(n-\delta-N-1)\\
\qquad{} = {\hat{d}}(x) d(x)- \Lambda ( x).
\end{gather*}
Since the left hand side must be independent of~$n$, option \eqref{a1} is ruled out.
Also option \eqref{a2} is ruled out: using \eqref{a2} and $\delta+N+1=0$ (from~\eqref{a2'}), the left hand side again cannot be independent of~$n$.
Only~\eqref{a3} remains, giving
\begin{gather*}
(n-\delta-N) (n-N)\,\frac{c_a}{ c_b} + n (n+\gamma) \frac{c_b}{ c_a} + (n+\gamma+1)(n-N)+
n(n-\delta-N-1)\\
\qquad{} = {\hat{d}}(x) d(x)- \Lambda ( x).
\end{gather*}
In order for $n^2$ in the left hand side to vanish, we again require ${c_a}/{ c_b}=-1$. This gives
\begin{gather*}
-N(N+\gamma+\delta+1) = {\hat{d}}(x) d(x)- \Lambda ( x),
\end{gather*}
and we see that both sides are indeed independent of~$n$.

The combined options \eqref{b2} and \eqref{a3} also give a valid set of equations of the form~\eqref{yyh} and~\eqref{hhy}, now corresponding to the parameter values
\begin{gather*}
\hat{\gamma} = \gamma, \qquad \hat{\delta}= \delta,\qquad \hat{N} = N-1.
\end{gather*}
Moreover, by means of \eqref{ab} we f\/ind $\xi=0$ and so $\hat{x} = x$.
Putting again $n=0$ in~\eqref{yyh} and~\eqref{hhy} for these $a$, ${\hat{a}}$, $b$, ${\hat{b}}$ we f\/ind
\begin{gather*}
\begin{split}
&(-\delta-N) R_0(\lambda(x);\gamma,\delta,N) - (\gamma+1) R_1(\lambda(x);\gamma,\delta,N) \\
&\qquad{}=
- \frac{(N-x)(x+\gamma+\delta+N+1)}{N} = \hat{d}(x)
\end{split}
\end{gather*}	
and similarly $d(x)= N$.
The relations in question are then, for $R_n(x)\equiv R_n(\lambda(x);\gamma,\delta,N)$ and $\hat R_n(x)\equiv R_n(\lambda(x);\gamma,\delta,N-1)$
\begin{gather*}
 (n-\delta-N) R_n(x) - (n+\gamma+1) R_{n+1}(x) =- \frac{(N-x)(x+\gamma+\delta+N+1)}{N} \hat R_{n}(x), \\
 (n+1) \hat R_{n}(x) -(n-N+1) \hat R_{n+1}(x) = N R_{n+1}(x),
\end{gather*}
which can be verif\/ied algebraically or by means of a computer algebra package.

 {\em Case \eqref{b3}}.
The next option to consider is~\eqref{b3}, for which \eqref{BD2} becomes
\begin{gather*}
a(n) (n+\gamma+1) c_b^{-1} + \hat{a}(n-1) (n-N-1) c_b + (n+\gamma+1)(n-N)+
n(n-\delta-N-1)\\
\qquad{} = {\hat{d}}(x) d(x)- \Lambda ( x).
\end{gather*}
The independence of $n$ in the left hand side again rules out options~\eqref{a1} and \eqref{a2}, while \eqref{a3} gives
\begin{gather*}
(n-\delta-N) (n+\gamma+1) \frac{c_a}{ c_b} + n (n-N-1)\frac{c_b}{ c_a} + (n+\gamma+1)(n-N)+
n(n-\delta-N-1)\\
\qquad{} = {\hat{d}}(x) d(x)- \Lambda ( x).
\end{gather*}
Also here, we require ${c_a}/{ c_b}=-1$ to arrive at a left hand side independent of~$n$, namely
\begin{gather*}
(\gamma+1)\delta = {\hat{d}}(x) d(x)- \Lambda ( x).
\end{gather*}
The combined options \eqref{b3} and \eqref{a3} thus give a valid set of equations of the form~\eqref{yyh} and~\eqref{hhy}, and they correspond to the parameter values
\begin{gather*}
\hat{\gamma} = \gamma+1, \qquad \hat{\delta}= \delta-1,\qquad \hat{N} = N;
\end{gather*}
by means of \eqref{ab} we f\/ind $\xi=0$ and so $\hat{x} = x$.
Finally, plugging these $a$, ${\hat{a}}$, $b$, ${\hat{b}}$ in~\eqref{yyh} and~\eqref{hhy} and putting $n=0$ we f\/ind
\begin{gather*}
(-\delta-N)R_{0}(\lambda(x);\gamma,\delta,N)+N R_{1}(\lambda(x);\gamma,\delta,N) =- \frac{(x+\gamma+1)(x+\delta)}{(\gamma+1)} = \hat{d}(x)
\end{gather*}	
and similarly $d(x)= \gamma+1$.

Hence we have the relations, for $R_n(x)\equiv R_n(\lambda(x);\gamma,\delta,N)$ and $\hat R_n(x)\equiv R_n(\lambda(x);\gamma+1$, $\delta-1,N)$
\begin{gather*}
 -(n-\delta-N)R_{n}(x)+(n-N) R_{n+1}(x) =\frac{(x+\gamma+1)(x+\delta)}{(\gamma+1)} \hat R_n(x),\\
 -(n+1) \hat R_n(x)+ (n+\gamma+2) \hat R_{n+1}(x) = (\gamma+1) R_{n+1}(x).
\end{gather*}
These can again be verif\/ied algebraically or by means of a computer algebra package.
Note that these relations coincide with~\eqref{R-rec1},~\eqref{R-rec2} from the previous section (up to a shift $\delta\rightarrow\delta+1$).

 {\em Case \eqref{b4}}.
The f\/inal option \eqref{b4} for $b$, $\hat{b}$ does not correspond to a valid set of equations of the form \eqref{yyh} and \eqref{hhy} as the left hand side of \eqref{BD2} can never be independent of $n$ for either options \eqref{a1}, \eqref{a2} or \eqref{a3}.

This completes the analysis in the case of dual Hahn polynomials, and we have the following result

\begin{Theorem}	\label{theo2}
	The only way to double dual Hahn polynomials, i.e., to combine two sets of dual Hahn polynomials such that they satisfy a pair of recurrence relations of the form~\eqref{rel1},~\eqref{rel2} is one of the three cases:

\noindent
{\bf dual Hahn~I},
	$R_n(x)\equiv R_n(\lambda(x);\gamma,\delta,N)$ and $\hat R_n(x)\equiv R_n(\lambda(x-1);\gamma+1,\delta+1,N-1)$:
	\begin{gather*}
	 R_{n}(x)- R_{n+1}(x) = \frac{x(x+\gamma+\delta+1)}{N(\gamma+1)} \hat R_{n}(x), \\
	 -(n+1)(N-n+\delta) \hat R_{n}(x)+(N-n-1)(n+\gamma+2) \hat R_{n+1}(x) = N(\gamma+1)R_{n+1}(x).
	\end{gather*}
 {\bf dual Hahn~II},
	$R_n(x)\equiv R_n(\lambda(x);\gamma,\delta,N)$ and $\hat R_n(x)\equiv R_n(\lambda(x);\gamma,\delta,N-1)$:
	\begin{gather*}
	 (n-\delta-N) R_n(x) - (n+\gamma+1) R_{n+1}(x) =- \frac{(N-x)(x+\gamma+\delta+N+1)}{N} \hat R_{n}(x), \\
	 (n+1) \hat R_{n}(x) -(n-N+1) \hat R_{n+1}(x) = N R_{n+1}(x).
	\end{gather*}
{\bf dual Hahn~III},
	$R_n(x)\equiv R_n(\lambda(x);\gamma,\delta,N)$ and $\hat R_n(x)\equiv R_n(\lambda(x);\gamma+1,\delta-1,N)$:
	\begin{gather*}
	 -(n-\delta-N)R_{n}(x)+(n-N) R_{n+1}(x) =\frac{(x+\gamma+1)(x+\delta)}{(\gamma+1)} \hat R_n(x),\\
	 -(n+1) \hat R_n(x)+ (n+\gamma+2) \hat R_{n+1}(x) = (\gamma+1) R_{n+1}(x).
	\end{gather*}
\end{Theorem}

By interchanging~$x$ and~$n$, each of the recurrence relations for dual Hahn polynomials in the previous theorem gives rise to a set of forward and backward shift operators for regular Hahn polynomials. The case {\bf dual Hahn~I} corresponds just to the known forward and backward shift operators for Hahn polynomials~\cite{Koekoek}: $Q_n(x)\equiv Q_n(x;\alpha,\beta,N)$ and $\hat Q_n(x)\equiv Q_n(x;\alpha+1,\beta+1$, $N-1)$:
\begin{gather*}
 Q_{n}(x)- Q_{n}(x+1) = \frac{n(n+\alpha+\beta+1)}{N(\alpha+1)} \hat Q_{n-1}(x), \\
 -(x+1)(N-x+\beta) \hat Q_{n-1}(x)+(N-x-1)(x+\alpha+2) \hat Q_{n-1}(x+1) \\
 \qquad{} = N(\alpha+1)Q_{n}(x+1).
\end{gather*}
The case {\bf dual Hahn~III} corresponds to our introductory example \eqref{Q-rec1},~\eqref{Q-rec2} (up to a~shift $\beta\rightarrow\beta+1$), and appears already in~\cite{Stoilova2011}.
The case {\bf dual Hahn~II} yields a new set of relations (encountered recently in~\cite[equations~(16),~(17)]{JSV2014}),
namely $Q_n(x)\equiv Q_n(x;\alpha,\beta,N)$ and $\hat Q_n(x)\equiv Q_n(x;\alpha,\beta,N-1)$:
\begin{gather*}
 (x-\beta-N) Q_n(x) - (x+\alpha+1) Q_n(x+1) =- \frac{(N-n)(n+\alpha+\beta+N+1)}{N} \hat Q_{n}(x), \\
 (x+1) \hat Q_{n}(x) -(x-N+1) \hat Q_{n}(x+1) = N Q_n(x+1).
\end{gather*}
The most important thing is, however, that we have classif\/ied the possible cases.

Because the sets of recurrence relations are of the form~\eqref{rel1},~\eqref{rel2}, they can be cast in matrix form, like in \eqref{MUUD}, with a simple two-diagonal matrix.
For the case {\bf dual Hahn~I}, note that the $N$-values of $R_n(x)$ and $\hat R_n(x)$ dif\/fer by~1, so the def\/inition of the matrix $U$ (again in terms of the normalized version of the polynomials) requires a little bit more attention.
The matrix $U$ is now of order $(2N+1)\times (2N+1)$ with matrix elements
\begin{gather}
 U_{2n,N-x} = U_{2n,N+x} = \frac{(-1)^n}{\sqrt{2}} \tilde R_n(\lambda(x);\gamma,\delta,N), \qquad x=1,\ldots,N, \nonumber \\
 U_{2n+1,N-x} = -U_{2n+1,N+x} = -\frac{(-1)^n}{\sqrt{2}} \tilde R_n(\lambda(x-1);\gamma+1,\delta+1,N-1), \qquad x=1,\ldots,N, \nonumber \\
 U_{2n,N}= (-1)^n \tilde R_n(\lambda(0);\gamma,\delta,N),\qquad U_{2n+1,N}=0,\label{U-HahnI}
\end{gather}
where the row index of the matrix $U$ (denoted here by $2n$ or $2n+1$, depending on the parity of the index) also runs over the integers from 0 up to $2N$.
This matrix $U$ is orthogonal: the orthogonality relation of the dual Hahn polynomials~\eqref{orth-R} and the signs in the matrix $U$ imply that its rows are orthonormal. Thus $U^TU=UU^T=I$, the identity matrix.
Then the recurrence relations for {\bf dual Hahn~I} of Theorem~\ref{theo2} are now reformulated in terms of a two-diagonal $(2N+1)\times(2N+1)$-matrix of the form
\begin{gather}
\label{MK-I}
M= \left( \begin{matrix}
 0 & M_0 & 0 & & \\
 M_0 & 0 & M_1 & \ddots & \\
 0 & M_1 & 0 & \ddots & 0 \\
 &\ddots & \ddots & \ddots & M_{2N-1} \\
 & & 0 & M_{2N-1} & 0
 \end{matrix} \right).
\end{gather}

Explicitly
\begin{Proposition}[dual Hahn~I]
\label{prop3}
Suppose $\gamma>-1$, $\delta>-1$.
Let $M$ be the two-diagonal mat\-rix~\eqref{MK-I} with
\begin{gather}
M_{2k}=\sqrt{(k+\gamma+1)(N-k)}, \qquad M_{2k+1}=\sqrt{(k+1)(N+\delta-k)},
\label{M_kI}
\end{gather}
and $U$ the orthogonal matrix determined in~\eqref{U-HahnI}.
Then the columns of $U$ are the eigenvectors of $M$, i.e., $M U = U D$,
where $D$ is a diagonal matrix containing the eigenvalues of~$M$
\begin{gather}
 D= \operatorname{diag} (-\epsilon_N,\ldots,-\epsilon_1,0,\epsilon_1,\ldots,\epsilon_{N}),\nonumber\\
 \epsilon_{k}=\sqrt{k(k+\gamma+\delta+1)},\qquad k=1,\ldots,N. \label{324}
\end{gather}
\end{Proposition}

Note that we have kept only the conditions under which the matrix $M$ is real.
The other conditions for which the dual Hahn polynomials in~\eqref{U-HahnI} can be normalized (namely $\gamma<-N$, $\delta<-N$)
would give rise to imaginary values in~\eqref{M_kI}.
In such a case, the relation $MU=UD$ remains valid, and also $D$ would have imaginary values.

For the case {\bf dual Hahn~II}, the matrix $U$ is again of order $(2N+1)\times (2N+1)$ with matrix elements
\begin{gather}
 U_{2n,x} = U_{2n,2N-x} = \frac{1}{\sqrt{2}} \tilde R_n(\lambda(x);\gamma,\delta,N), \qquad x=0,\ldots,N-1, \nonumber \\
 U_{2n+1,x} = -U_{2n+1,2N-x} = -\frac{1}{\sqrt{2}} \tilde R_n(\lambda(x);\gamma,\delta,N-1), \qquad x=0,\ldots,N-1, \nonumber \\
 U_{2n,N}= \tilde R_n(\lambda(N);\gamma,\delta,N),\qquad U_{2n+1,N}=0,\label{U-HahnII}
\end{gather}
where the row indices are as in~\eqref{U-HahnI}.
The orthogonality relation of the dual Hahn polynomials and the signs in the matrix $U$ imply that its rows are orthonormal, so $U^TU=UU^T=I$.
The pair of recurrence relations for {\bf dual Hahn~II} of Theorem~\ref{theo2} yield
\begin{Proposition}[dual Hahn~II]\label{prop4}
Suppose $\gamma>-1$, $\delta>-1$.
Let $M$ be a tridiagonal $(2N+1)\times(2N+1)$-matrix of the form~\eqref{MK-I} with
\begin{gather}
M_{2k}= \sqrt{(N+\delta-k)(N-k)}, \qquad M_{2k+1}=\sqrt{(k+1)(k+\gamma+1)},
\label{M_kII}
\end{gather}
and $U$ the orthogonal matrix determined in~\eqref{U-HahnII}.
Then the columns of $U$ are the eigenvectors of~$M$, i.e., $M U = U D$,
where~$D$ is a diagonal matrix containing the eigenvalues of~$M$
\begin{gather*}
D= \operatorname{diag} (-\epsilon_N,\ldots,-\epsilon_1,0,\epsilon_1,\ldots,\epsilon_{N}), \\
 \epsilon_{k}=\sqrt{k(\gamma+\delta+1+2N-k)},\qquad k=1,\ldots,N.
\end{gather*}
\end{Proposition}

Note that the order in which the normalized dual Hahn polynomials appear in the matrix $U$ is dif\/ferent for~\eqref{U-HahnI} and~\eqref{U-HahnII}.
This is related to the indices of the polynomials in the relations of Theorem~\ref{theo2}.

Finally, for the case {\bf dual Hahn~III}, the matrix $U$ is given by~\eqref{UevenR},~\eqref{UoddR} and we recapitulate the results given at the end of the previous section, now in terms of the dual Hahn parameters~$\gamma$ and~$\delta$.
\begin{Proposition}[dual Hahn~III]
	\label{prop1}
	Suppose $\gamma>-1$, $\delta>-1$ or $\gamma<-N-1$, $\delta<-N-1$.
	Let $M$ be the tridiagonal mat\-rix~\eqref{MK} with
	\begin{gather}
	M_{2k}= \sqrt{(k+\gamma+1)(N+\delta+1-k)}, \qquad M_{2k+1}=\sqrt{(k+1)(N-k)},
	\label{M_kIII}
	\end{gather}
	and $U$ the orthogonal matrix determined in~\eqref{UevenR},~\eqref{UoddR}.
	Then the columns of $U$ are the eigenvectors of $M$, i.e., $M U = U D$,
	where $D$ is a diagonal matrix containing the eigenvalues of~$M$
	\begin{gather*}
 D= \operatorname{diag} (-\epsilon_N,\ldots,-\epsilon_1,-\epsilon_0,\epsilon_0,\epsilon_1,\ldots,\epsilon_{N}),\\
	\epsilon_{k}=\sqrt{(k+\gamma+1)(k+\delta+1)}, \qquad
 k=0,1,\ldots,N.
	\end{gather*}
\end{Proposition}

To conclude for dual Hahn polynomials: there are three sets of recurrence relations of the form~\eqref{rel1},~\eqref{rel2}. Each of the three cases gives rise to a two-diagonal matrix with simple and explicit eigenvalues, and eigenvectors given in terms of two sets of dual Hahn polynomials.

\section{Doubling Hahn polynomials} \label{sec:Hahn}

The technique presented in the previous section can be applied to other types of discrete ortho\-go\-nal polynomials with a f\/inite spectrum. We have done this for Hahn polynomials.
One level up in the hierarchy of orthogonal polynomials of hypergeometric type are the Racah polynomials. Also for Racah polynomials we have applied the technique, but here the description of the results becomes very technical.
So we shall leave the results for Racah polynomials for Appendix~\ref{sec:Racah}.

For Hahn polynomials the analysis is again straightforward but tedious, so let us skip the details of the computation and present just the f\/inal outcome here.
Applying the technique described in \eqref{yyh}--\eqref{LL}, with $y_n = Q_n(x;\alpha,\beta,N)$ and
$\hat{y}_n = Q_{n}(\hat{x};\hat{\alpha},\hat{\beta},\hat{N})$ yields the following result.

\begin{Theorem}
\label{theo5}
The only way to combine two sets of Hahn polynomials such that they satisfy a~pair of recurrence relations
of the form~\eqref{yyh},~\eqref{hhy} is one of the four cases:\\
{\bf Hahn~I},
$Q_n(x)\equiv Q_n(x;\alpha,\beta,N)$ and $\hat Q_n(x)\equiv Q_n(x;\alpha+1,\beta,N)$:
\begin{gather*}
 \frac{(n+\alpha+\beta+N+2)}{(2n+\alpha+\beta+2)} Q_n(x) - \frac{(N-n)}{(2n+\alpha+\beta+2)} Q_{n+1}(x)
	= \frac{(\alpha+x+1)}{(\alpha+1)} \hat Q_n(x), \\
 - \frac{(n+1)(n+\beta+1) }{(2n+\alpha+\beta+3)} \hat Q_n(x) + \frac{(n+\alpha+\beta+2)(n+\alpha+2)}{(2n+\alpha+\beta+3)} \hat Q_{n+1}(x)
 = (\alpha+1) Q_{n+1}(x).
\end{gather*}	
{\bf Hahn~II},
$Q_n(x)\equiv Q_n(x;\alpha,\beta,N)$ and $\hat Q_n(x)\equiv Q_n(x-1;\alpha+1,\beta,N-1)$:
\begin{gather*}
 \frac{1}{(2n+\alpha+\beta+2)} Q_n(x) - \frac{1}{(2n+\alpha+\beta+2)} Q_{n+1}(x)
	= \frac{x}{N(\alpha+1)} \hat Q_{n}(x), \\
 - \frac{(n+1)(n+\beta+1) (n+\alpha+\beta+N+2)}{(2n+\alpha+\beta+3)} \hat Q_{n}(x) \\
 \qquad{} + \frac{(n+\alpha+\beta+2)(N-n-1)(n+\alpha+2)}{(2n+\alpha+\beta+3)} \hat Q_{n+1}(x)
	= N(\alpha+1) Q_{n+1}(x) .
\end{gather*}	
{\bf Hahn~III},
$Q_n(x)\equiv Q_n(x;\alpha,\beta,N)$ and $\hat Q_n(x)\equiv Q_n(x;\alpha,\beta+1,N)$:
\begin{gather*}
 \frac{(n+\beta+1 )(n+N+2+\alpha+\beta)}{(2n+\alpha+\beta+2)} Q_n(x) +\frac{(N-n)(n+\alpha+1)}{(2n+\alpha+\beta+2)} Q_{n+1}(x)\\
 \qquad{}	= (\beta+1+N-x) \hat Q_n(x), \\
 \frac{(n+1)}{(2n+\alpha+\beta+3)} \hat Q_n(x) + \frac{(n+\alpha+\beta+2)}{(2n+\alpha+\beta+3)} \hat Q_{n+1}(x)
	= Q_{n+1}(x) .
\end{gather*}	
{\bf Hahn~IV},
$Q_n(x)\equiv Q_n(x;\alpha,\beta,N)$ and $\hat Q_n(x)\equiv Q_n(x;\alpha,\beta+1,N-1)$:
\begin{gather*}
 \frac{(n+\beta+1)}{(2n+\alpha+\beta+2)} Q_n(x) +\frac{(n+\alpha+1)}{(2n+\alpha+\beta+2)} Q_{n+1}(x)
	= \frac{(N-x)}{N} \hat Q_n(x) , \\
 \frac{(n+1)(n+\alpha+\beta+N+2)}{(2n+\alpha+\beta+3)} \hat Q_n(x) + \frac{(N-n-1)(n+\alpha+\beta+2)}{(2n+\alpha+\beta+3)} \hat Q_{n+1}(x)
	= N Q_{n+1}(x) .
\end{gather*}	
\end{Theorem}

Note that when interchanging $x$ and $n$ the relations in {\bf Hahn~II} coincide with the known forward and backward shift operator relations for dual Hahn polynomials~\cite[equations (9.6.6) and (9.6.8)]{Koekoek}.
In the same way, the other cases yield new forward and backward shift operator relations for dual Hahn polynomials.

Since the recurrence relations are of the form~\eqref{yyh},~\eqref{hhy}, they can be cast in matrix form with a two-diagonal matrix.
We shall write the matrix elements again in terms of normalized polynomials.
For the case {\bf Hahn~I}, the matrix $U$ of order $(2N+2)\times(2N+2)$, with elements
\begin{gather}
 U_{2n,N-x} = U_{2n,N+x+1} = \frac{(-1)^n}{\sqrt{2}} \tilde Q_n(x;\alpha,\beta,N), \nonumber\\
 U_{2n+1,N-x} = -U_{2n+1,N+x+1} = -\frac{(-1)^n}{\sqrt{2}} \tilde Q_n(x;\alpha+1,\beta,N) \label{dU-I}
\end{gather}
where $x,n\in\{0,1,\ldots,N\}$, is orthogonal, and the recurrence relations yield
\begin{Proposition}[Hahn~I]
\label{prop6}
Suppose that $\gamma,\delta >-1$.
Let $M$ be a tridiagonal $(2N+2)\times(2N+2)$-matrix of the form~\eqref{MK} with
\begin{gather*}
 M_{2k} = \sqrt{ \frac{	(k+\alpha+1)(k+\alpha+\beta+1)(k+\alpha+\beta+2+N)}{(2k+\alpha+\beta+1)(2k+\alpha+\beta+2)}}, \nonumber \\
 M_{2k+1} = \sqrt{ \frac{(k+\beta+1)(k+1)(N-k) }{(2k+\alpha+\beta+2)(2k+\alpha+\beta+3) } },
\end{gather*}
and $U$ the orthogonal matrix determined in~\eqref{dU-I}.
Then the columns of $U$ are the eigenvectors of $M$, i.e., $M U = U D$,
where $D$ is a diagonal matrix containing the eigenvalues of~$M$
\begin{gather}
D= \operatorname{diag} (-\epsilon_N,\ldots,-\epsilon_1,-\epsilon_0,\epsilon_0,\epsilon_1,\ldots,\epsilon_{N}),\nonumber\\
\epsilon_{k}=\sqrt{k+\alpha+1},\qquad k=0,1,\ldots,N.
\label{47}
\end{gather}
\end{Proposition}

For the case {\bf Hahn~II}, the orthogonal matrix $U$ is of order $(2N+1)\times(2N+1)$, with elements
\begin{gather}
 U_{2n,N-x} = U_{2n,N+x} = \frac{(-1)^n}{\sqrt{2}} \tilde Q_n(x;\alpha,\beta,N), \qquad x=1,\ldots,N, \nonumber \\
 U_{2n+1,N-x} = -U_{2n+1,N+x} = -\frac{(-1)^n}{\sqrt{2}} \tilde Q_n(x-1;\alpha+1,\beta,N-1), \qquad x=1,\ldots,N, \label{dU-II} \\
 U_{2n,N}= (-1)^n \tilde Q_n(0;\alpha,\beta,N),\qquad U_{2n+1,N}=0, \nonumber
\end{gather}
where the row indices are as in \eqref{U-HahnI}.
The recurrence relations for {\bf Hahn~II} yield
\begin{Proposition}[Hahn~II]
\label{prop7}
Suppose that $\alpha,\beta >-1$ or $\alpha,\beta<-N$.
Let $M$ be a tridiagonal $(2N+1)\times(2N+1)$-matrix of the form~\eqref{MK-I} with
\begin{gather*}
 M_{2k} = \sqrt{ \frac{	(k+\alpha+1)(k+\alpha+\beta+1)(N-k)}{(2k+\alpha+\beta+1)(2k+\alpha+\beta+2)}} ,\nonumber \\
 M_{2k+1} = \sqrt{ \frac{(k+\beta+1)(k+\alpha+\beta+2+N)(k+1) }{(2k+\alpha+\beta+2)(2k+\alpha+\beta+3) } },
\end{gather*}
and $U$ the orthogonal matrix determined in~\eqref{dU-II}.
Then the columns of $U$ are the eigenvectors of $M$, i.e., $M U = U D$,
where $D$ is a diagonal matrix containing the eigenvalues of~$M$:
\begin{gather*}
 D= \operatorname{diag} (-\epsilon_N,\ldots,-\epsilon_1,0,\epsilon_1,\ldots,\epsilon_{N}), \qquad
 \epsilon_{k}=\sqrt{k},\qquad k=1,\ldots,N.
\end{gather*}
\end{Proposition}

Note that for both cases, the two-diagonal matrix $M$ becomes more complicated compared to the cases for dual Hahn polynomials, but the matrix $D$ of eigenvalues becomes simpler.

For the two remaining cases we need not give all details: the matrix $M$ for the case {\bf Hahn~III} is equal to the matrix $M$ for
the case {\bf Hahn~I} with the replacement $\alpha \leftrightarrow \beta$, and so its eigenvalues are $\pm \sqrt{k+\beta+1}$, $k=0,1,\ldots,N$.
And the matrix~$M$ for the case {\bf Hahn~IV} is equal to the matrix~$M$ for
the case {\bf Hahn~II} with the same replacement $\alpha \leftrightarrow \beta$, so its eigenvalues are~$0$ and~$\pm \sqrt{k}$, $k=1,\ldots,N$.

\section{Polynomial systems, Christof\/fel and Geronimus transforms} \label{sec:orthpoly}

So far, we have only partially explained why the technique in the previous sections is referred to as ``doubling'' polynomials.
It is indeed a fact that the combination of two sets of polynomials, each with dif\/ferent parameters, yields a new set of orthogonal polynomials.
This can be compared to the well known situation of combining two sets of generalized Laguerre polynomials (both with dif\/ferent parameters $\alpha$ and $\alpha-1$) into the set of ``generalized Hermite polynomials''~\cite{Chihara}. There, for $\alpha>0$, one def\/ines
\begin{gather}
P_{2n}(x) = \sqrt{\frac{n!}{(\alpha)_n}} L_n^{(\alpha-1)}\big(x^2\big),\qquad
P_{2n+1}(x) =\sqrt{\frac{n!}{(\alpha)_{n+1}}} x L_n^{(\alpha)}\big(x^2\big). \label{Lcase2}
\end{gather}
Then the orthogonality relation of Laguerre polynomials leads to the orthogonality of the polynomials~\eqref{Lcase2}:
\begin{gather*}
\int_{-\infty}^{+\infty} w(x) P_n(x)P_{n'}(x)dx = \Gamma(\alpha) \delta_{n,n'}, 
\end{gather*}
where
\begin{gather}
w(x) = e^{-x^2} |x|^{2\alpha-1}.
\label{w2}
\end{gather}
Note that the even polynomials are Laguerre polynomials in $x^2$ (for parameter~$\alpha-1$), and the odd polynomials are Laguerre polynomials in~$x^2$ (for parameter $\alpha$) multiplied by a factor~$x$. The weight function~\eqref{w2} is common for both types of polynomials.
It is this phenomenon that appears here too in our doubling process of Hahn or dual Hahn polynomials.

From a more general point of view, this f\/its in the context of obtaining a new family of orthogonal polynomials starting from a set of
orthogonal polynomials and its kernel partner related by a Christof\/fel transform~\cite{Chihara,Marcellan,Vinet2012}.
In a way, our classif\/ication determines for which Christof\/fel parameter~$\nu$ (see~\cite{Vinet2012} for the notation) the Christof\/fel transform
of a Hahn, dual Hahn or Racah polynomial is again a Hahn, dual Hahn or Racah polynomial with possibly dif\/ferent parameters. This determines moreover quite explicitly the common weight function.

For a dual Hahn polynomial $R_n(x)\equiv R_n(\lambda(x);\gamma,\delta,N)$, with data given in~\eqref{AC},
and a~Chris\-tof\/fel parameter~$\nu$ the kernel partner is given by the transform
\begin{gather}\label{CT}
P_n(x) = \frac{R_{n+1}(x)-a_n R_n(x)}{\Lambda(x)-\Lambda(\nu)},\qquad a_n = \frac{R_{n+1}(\nu)}{R_n(\nu)}.
\end{gather}
Because of the recurrence relation~\eqref{re} and what is called the Geronimus transform the original polynomials can also be expressed in terms of the kernel partners.
This is usually done for monic polynomials (see~\cite[equations~(3.2) and~(3.3)]{Vinet2012}),
but it can be extended to non-monic dual Hahn polynomials as follows
\begin{gather}\label{GT}
R_n(x) = A(n) P_n(x) - b_n P_{n-1}(x)
\end{gather}
where the coef\/f\/icients $b_n$ are related to the recurrence relation~\eqref{re} as follows
\begin{gather}
b_na_{n-1} = C(n),\qquad A(n)a_n+b_n =A(n)+C(n)+\Lambda(\nu).
\label{baC}
\end{gather}

Our classif\/ication now shows that only for $\nu$ equal to one of the values $0$, $N$ or $-\delta$, the kernel partner $P_n(x)$ will again be a dual Hahn polynomial.
Indeed, taking for example $\nu =0$ in~\eqref{CT} we have $R_n(0) = 1$ and
\begin{gather*}
P_n(x)= \frac{R_{n+1}(x)-R_n(x)}{\Lambda(x)} = \frac{-1}{N(\gamma+1)}R_{n}(\lambda ( x-1);{\gamma+1},{\delta+1},{N-1}),
\end{gather*}
where we used the f\/irst relation of {\bf dual Hahn~I} to obtain again a dual Hahn polynomial.
The reverse transform~\eqref{GT} follows immediately from the second relation of {\bf dual Hahn~I}.
Similarly, taking $\nu=N$ in~\eqref{CT} we have $R_n(N) = {(-N-\delta)_{n}}/{(\gamma+1)_n}$ and
\begin{gather*}
P_n(x)
 = \frac{R_{n+1}(x)-R_n(x)(n-\delta-N)/(n+\gamma+1)}{(x-N)(x+N+\gamma+\delta+1)}
= \frac{-1}{N(n+\gamma+1)}R_n(\lambda(x);\gamma,\delta,N-1),
\end{gather*}
which we obtained using the f\/irst relation of {\bf dual Hahn~II}.
For the reverse transform~\eqref{GT} we f\/ind, using the second relation of {\bf dual Hahn~II} with shifted $n\mapsto n-1$,
\begin{gather*}
A(n) P_n(x) - b_n P_{n-1}(x) = \frac{-(n-N)}{N}R_n(\lambda(x);\gamma,\delta,N-1) + \frac{n}{N}R_{n-1}(\lambda(x);\gamma,\delta,N-1)\\
\hphantom{A(n) P_n(x) - b_n P_{n-1}(x)}{} = R_n(x).
\end{gather*}
For the last case, taking $\nu=-\delta$ in~\eqref{CT} we have $R_n(-\delta) = {(-N-\delta)_{n}}/{(-N)_n}$ and
\begin{gather*}
P_n(x) = \frac{R_{n+1}(x)-R_n(x)(n-\delta-N)/(n-N)}{(x+\gamma+1)(x+\delta) } \\
\hphantom{P_n(x)}{}
= \frac{1}{(\gamma+1)(n-N)}R_n(\lambda(x);\gamma+1,\delta-1,N),
\end{gather*}
which we obtained using the f\/irst relation of {\bf dual Hahn~III}. For the transform~\eqref{GT} we have
\begin{gather*}
A(n) P_n(x) - b_n P_{n-1}(x) = \frac{n+\gamma+1}{\gamma+1}R_n(\lambda(x);\gamma+1,\delta-1,N-1) \\
\hphantom{A(n) P_n(x) - b_n P_{n-1}(x) =}{} - \frac{n}{\gamma+1}R_{n-1}(\lambda(x);\gamma+1,\delta-1,N),
\end{gather*}
which equals $R_n(x)$ by the second relation of {\bf dual Hahn~III}.

In a similar way, for the Hahn polynomials, putting $Q_n(x)\equiv Q_n(x;a,b,N)$, using the data~\eqref{ABD} in
\begin{gather*}
P_n(x) = \frac{Q_{n+1}(x)-a_n Q_n(x)}{\Lambda(x)-\Lambda(\nu)},\qquad a_n = \frac{Q_{n+1}(\nu)}{Q_n(\nu)},
\end{gather*}
and in~\eqref{GT},~\eqref{baC},
the cases {\bf Hahn~I}, {\bf II}, {\bf III}, {\bf V} correspond respectively to the choices $-\alpha-1$, $0$, $N+\beta+1$ and $N$ for $\nu$.

The task of determining for which Christof\/fel parameter $\nu$ the kernel partner of a dual Hahn polynomial is again of the same family is not trivial. It comes down to f\/inding a pair of recurrence relations of the form~\eqref{rel1},~\eqref{rel2} with coef\/f\/icients related to $\nu$ as in~\eqref{CT}. We have classif\/ied these for general coef\/f\/icients, without a relation to~$\nu$, and we observe that each solution indeed corresponds to a specif\/ic choice for~$\nu$.

The transforms~\eqref{CT}, \eqref{GT} give rise to new orthogonal systems, but in general there is no way of writing the common weight function. However, since here both sets are of the same family, we can actually do this.
Let us begin with the dual Hahn polynomials, in particular the case {\bf dual Hahn~I}, for which the corresponding matrix~$U$ is given in~\eqref{U-HahnI}. They give rise to a new family of discrete orthogonal polynomials with the relation $MU=UD$ corresponding to their three term recurrence relation with Jacobi matrix~$M$~\eqref{M_kI}. In general the support of the weight function is equal to the spectrum of the Jacobi matrix~\cite{Berezanskii,Klimyk2006,Koelink2004,Koelink1998}.
After simplifying with the normalization factors~\eqref{R-tilde}, this leads to a discrete orthogonality of polynomials, with support equal
to the eigenvalues of $M$ (so in this case, the support follows from~\eqref{324}).
Concretely, for the case under consideration, we have
\begin{Proposition}
	\label{prop10}
	Let $\gamma>-1$, $\delta>-1$, and consider the $2N+1$ polynomials
	\begin{gather*}
	{P}_{2n} (q) = \frac{(-1)^n}{\sqrt{2}} {R}_n\big(q^2;\gamma,\delta,N\big), \qquad n=0,1,\ldots,N,\nonumber\\
	P_{2n+1} (q) = - \frac{(-1)^n}{\sqrt{2}}\frac{\sqrt{(n+\gamma+1)(N-n)}}{(\gamma+1)N} q {R}_n\big(q^2-\gamma-\delta-2;\gamma+1,\delta+1,N-1\big),\nonumber\\
\hphantom{P_{2n+1} (q) =}{} n=0,1,\ldots,N-1.
	\end{gather*}
	These polynomials satisfy the discrete orthogonality relation
	\begin{gather}
	 \sum_{q\in S} \frac{(-1)^k (2k+\gamma +\delta +1) (\gamma+1)_k (-N)_k N! }{ (k+\gamma +\delta +1)_{N+1} (\delta+1)_k k! }
	(1+\delta_{q,0}) P_n(q) P_{n'}(q) \nonumber\\
	 \qquad {}= \left[ \binom{\gamma+\lfloor n/2\rfloor }{\lfloor n/2\rfloor} \binom{\delta+N-\lfloor n/2\rfloor }{N-\lfloor n/2\rfloor} \right]^{-1}
	\delta_{n,n'}
	\label{511}
	\end{gather}
	with
	\begin{gather*}
	S=\big\{ 0,\pm\sqrt{k(k+\gamma+\delta+1)}, \ k=1,2,\ldots,N\big\}.
	\end{gather*}
\end{Proposition}

Note that for $q\in S$, $q^2=k(k+\gamma+\delta+1)\equiv \lambda(k)$, and the polynomial $P_{2n}(q)$ is of the form $R_n(\lambda(k);\gamma,\delta,N)$.
In that case, $q^2-\gamma-\delta-2 = (k-1)((k-1)+(\gamma+1)+(\delta+1)+1) \equiv \lambda(k-1)$, so the polynomial $P_{2n+1}(q)$ is of the form $R_n(\hat \lambda(k-1);\gamma+1,\delta+1,N-1)$.
The interpretation of the weight function in the left hand side of~\eqref{511} is as follows: each~$q$ in the support~$S$ is mapped to a $k$-value belonging to $\{0,1,\ldots,N\}$, and then the weight depends on this $k$-value.

Now we turn to the classif\/ication of Section~\ref{sec:Hahn}, where the corresponding orthogonal mat\-ri\-ces~$U$ are given in terms of (normalized) Hahn polynomials. for the case {\bf Hahn~I}, the matrix~$U$ is given in~\eqref{dU-I}, and the spectrum of the matrix~$M$ is given by~\eqref{47}.
After simplifying the normalization factors, the orthogonality of the rows of $U$ gives rise to
\begin{Proposition}
\label{prop8}
Let $\alpha>-1$, $\beta>-1$, and consider the $2N+2$ polynomials, $n=0,1,\ldots,N$,
\begin{gather*}
 	{P}_{2n} (q) = \frac{(-1)^n}{\sqrt{2}} {Q}_n\big(q^2-\alpha-1;\alpha,\beta,N\big),\nonumber\\
 	P_{2n+1} (q) = - \frac{(-1)^n}{\sqrt{2}}\frac{1}{(\alpha+1)} \sqrt{\frac{(n+\alpha+1)(n+\alpha+\beta+1)(2n+2+\alpha+\beta)}{(n+N+\alpha+\beta+2)(2n+\alpha+\beta+1)}}\nonumber\\
\hphantom{P_{2n+1} (q) =}{} \times q {Q}_n\big(q^2-\alpha-1;\alpha+1,\beta,N\big).
\end{gather*}
These polynomials satisfy the discrete orthogonality relation
\begin{gather*}
	\sum_{q\in S} \binom{ q^2-1}{q^2-\alpha-1} \binom{ N - q^2+\alpha+\beta +1}{N - q^2+\alpha+1} P_n(q) P_{n'}(q) =h_{\lfloor n/2\rfloor}(\alpha,\beta,N) \beta_{n,n'}
\end{gather*}
with
\begin{gather*}
S=\big\{ {-}\sqrt{N+\alpha+1}, -\sqrt{N+\alpha}, \ldots, -\sqrt{\alpha+1}, \sqrt{\alpha+1}, \ldots, \sqrt{N+\alpha},\sqrt{N+\alpha+1}\big\}
\end{gather*}
and
\begin{gather*}
h_{n}(\alpha,\beta,N) = \frac{(-1)^n (n+\alpha +\beta +1)_{N+1} (\beta+1)_n n! }{ (2n+\alpha +\beta +1) (\alpha+1)_n (-N)_n N! }.
\end{gather*}
\end{Proposition}

So $P_n(q)$ is a polynomial of degree $n$ in the variable~$q$, of dif\/ferent type (with dif\/ferent parameters when expressed as a Hahn polynomial) depending on whether~$n$ is even or~$n$ is odd. The support points of the discrete orthogonality are given by
\begin{gather*}
q = \pm\sqrt{k+ \alpha+1}, \qquad k=0,\dots, N.
\end{gather*}

In the same way, the dual orthogonality for the case {\bf Hahn~II} gives rise to
\begin{Proposition}
\label{prop9}
Let $\alpha>-1$, $\beta>-1$, and consider the $2N+1$ polynomials
\begin{gather*}
{P}_{2n} (q) = \frac{(-1)^n}{\sqrt{2}} {Q}_n\big(q^2;\alpha,\beta,N\big), \qquad n=0,1,\ldots,N,\nonumber\\
P_{2n+1} (q) = - \frac{(-1)^n}{\sqrt{2}}\frac{1}{(\alpha+1)N} \sqrt{\frac{(N-n)(n+\alpha+1)(n+\alpha+\beta+1)(2n+\alpha+\beta+2)}
{(2n+\alpha+\beta+1)}} \nonumber\\
\hphantom{P_{2n+1} (q) =}{} \times q {Q}_n\big(q^2-1;\alpha+1,\beta,N-1\big), \qquad n=0,1,\ldots,N-1.
\end{gather*}
These polynomials satisfy the discrete orthogonality relation
\begin{gather*}
\sum_{q\in S} \binom{q^2+\alpha}{q^2} \binom{N - q^2+\beta}{N - q^2}
(1+\delta_{q,0}) P_n(q) P_{n'}(q) =h_{\lfloor n/2\rfloor}(\alpha,\beta,N) \beta_{n,n'}
\end{gather*}
with
\begin{gather*}
S=\big\{ {-}\sqrt{N}, -\sqrt{N-1}, \ldots, -1,0,1,\ldots, \sqrt{N-1},\sqrt{N}\big\}
\end{gather*}
and
\begin{gather*}
h_{n}(\alpha,\beta,N) = \frac{(-1)^n (n+\alpha +\beta +1)_{N+1} (\beta+1)_n n! }{ (2n+\alpha +\beta +1) (\alpha+1)_n (-N)_n N! }.
\end{gather*}
\end{Proposition}

The ideas described in the three propositions of this section should be clear.
It would lead us too far to give also the explicit forms corresponding to the remaining cases.
Let us just mention that also for these cases the support of the new polynomials coincides with the spectrum of the corresponding two-diagonal matrix $M$.

\section{First application: eigenvalue test matrices} \label{sec:testmatrix}

In Sections~\ref{sec:dHahn} and \ref{sec:Hahn} we have encountered a number of symmetric two-diagonal matrices $M$ with explicit expressions for the eigenvectors and eigenvalues.
In general, if one considers a~two-diagonal matrix~$A$ of size $(m+2)\times(m+2)$,
\begin{gather}
\label{A}
A= \left( \begin{matrix}
 0 & b_0 & 0 & & \\
 c_0 & 0 & b_1 & \ddots & \\
 0 & c_1 & 0 & \ddots & 0 \\
 &\ddots & \ddots & \ddots & b_{m} \\
 & & 0 & c_{m} & 0
 \end{matrix} \right),
\end{gather}
then it is clear that the characteristic polynomial depends on the products $b_ic_i$, $i=0,\ldots,m$, only, and not on~$b_i$ and $c_i$ separately.
So the same holds for the eigenvalues.
Therefore, if all matrix elements $b_i$ and $c_i$ are positive, the eigenvalues of~$A$ or of the related symmetric matrix
\begin{gather*}
A'= \left( \begin{matrix}
 0 & \sqrt{b_0c_0} & 0 & & \\
 \sqrt{b_0c_0} & 0 & \sqrt{b_1c_1} & \ddots & \\
 0 & \sqrt{b_1c_1} & 0 & \ddots & 0 \\
 &\ddots & \ddots & \ddots & \sqrt{b_{m}c_m} \\
 & & 0 & \sqrt{b_{m}c_m} & 0
 \end{matrix} \right)
\end{gather*}
are the same. The eigenvectors of $A'$ are those of $A$ after multiplication by a diagonal matrix (the diagonal matrix that is used
in the similarity transformation from~$A$ to~$A'$).

For matrices of type~\eqref{A}, it is suf\/f\/icient to denote them by their superdiagonal $[\mathbf{b}]=[b_0,\ldots,b_m]$
and their subdiagonal $[\mathbf{c}]=[c_0,\ldots,c_m]$. So the Sylvester--Kac matrix from the introduction is denoted by
\begin{gather*}
[\mathbf{b}]=[1,2,\ldots,N],\qquad [\mathbf{c}]=[N,\ldots,2,1],
\end{gather*}
with eigenvalues given by~\eqref{KacEig}.

The importance of the Sylvester--Kac matrix as a test matrix for numerical eigenvalue routines has already been emphasized in the Introduction.
In this context, it is also signif\/icant that the matrix itself has integer entries only (so there is no rounding error when represented on a digital computer), and that also the eigenvalues are integers. Of course, matrices with rational numbers as entries suf\/f\/ice as well, since one can always multiply the matrix by an appropriate integer factor.

Let us now systematically consider the two-diagonal matrices encountered in the classif\/ication process of doubling Hahn or dual Hahn polynomials.
For the matrix~\eqref{MK-I} of the {\bf dual Hahn~I} case, the corresponding non-symmetric form can be chosen as the two-diagonal matrix with
\begin{gather}
[\mathbf{b}]=[\gamma+1,1,\gamma+2,2,\ldots,\gamma+N,N], \nonumber\\
[\mathbf{c}]=[N,N+\delta,N-1,N-1+\delta,\ldots,1,\delta+1].
\label{Mat1}
\end{gather}
The eigenvalues are determined by Proposition~\ref{prop3} and given by~$0$, $\pm\sqrt{k(k+\gamma+\delta+1)}$, $k=1,\ldots,N$.
This is (up to a factor~2) the matrix~\eqref{Kac-odd} mentioned in the Introduction.
As test matrix, the choice $\gamma+\delta+1=0$ (leaving one free parameter) is interesting as it gives rise to integer eigenvalues.
In Proposition~\ref{prop3} there is the initial condition $\gamma>-1$, $\delta>-1$.
Clearly, if one is only dealing with eigenvalues, the condition for~\eqref{Mat1} is just $\gamma+\delta+2\geq 0$.
And when one substitutes $\delta=-\gamma-1$ in~\eqref{Mat1}, there is no condition at all for the one-parameter family of matrices
of the form~\eqref{Mat1}.

For the {\bf dual Hahn~II} case, the matrix~\eqref{M_kII} is given in Proposition~\ref{prop4}, and its non-symmetric form can be
taken as
\begin{gather}
[\mathbf{b}]=[\gamma+N,1,\gamma+N-1,2,\ldots,\gamma+1,N], \nonumber\\
[\mathbf{c}]=[N,\delta+1,N-1,\delta+2,\ldots,1,\delta+N].
\label{Mat2}
\end{gather}
The eigenvalues are given by $0$, $\pm\sqrt{k(\gamma+\delta+1+2N-k)}$, $k=1,\ldots,N$.
There is no simple substitution that reduces these eigenvalues to integers.

For the {\bf dual Hahn~III} case, the matrix~\eqref{M_k} is given in Proposition~\ref{prop1}, and its simplest non-symmetric form
is
\begin{gather}
[\mathbf{b}]=[\gamma+1,1,\gamma+2,2,\ldots,\gamma+N,N,\gamma+N+1], \nonumber\\
[\mathbf{c}]=[\delta+N+1,N,\delta+N,N-1,\ldots,\delta+2,1,\delta+1].
\label{Mat3}
\end{gather}
The eigenvalues are given by~\eqref{epsilon}, i.e., $\pm\sqrt{(\gamma+k+1)(\delta+k+1)}$, $k=0,\ldots,N$.
Up to a~fac\-tor~2, this is the third matrix mentioned in the Introduction.
The substitution $\delta=\gamma$ leads to a~one-parameter family of two-diagonal matrices with square-free eigenvalues.
And in particular when moreover $\gamma$ is integer, all matrix entries and all eigenvalues are integers.

The two-diagonal matrices arising from the Hahn doubles or the Racah doubles can also be written in a square-free form of type~\eqref{A}.
However, for these cases the entries in the two-diagonal matrices $M$ are already quite involved (see, e.g., Propositions~\ref{prop6}, \ref{prop7}, \ref{propR1} or \ref{propR2}), and we shall not discuss them further in this context.
The three examples given here, \eqref{Mat1}--\eqref{Mat3}, are already suf\/f\/iciently interesting as extensions of the Sylvester--Kac matrix as potential eigenvalue test matrices.

\section[Further applications: related algebraic structures and f\/inite oscillator models]{Further applications: related algebraic structures\\ and f\/inite oscillator models} \label{sec:oscillators}

The original example of a (dual) Hahn double, described here in Section~\ref{sec:Example},
was encountered in the context of a f\/inite oscillator model~\cite{JSV2011}.
In that context, there is also a related algebraic structure.
In particular, the two-diagonal matrices $M$ of the form~\eqref{MK} or~\eqref{MK-I} are interpreted as representation matrices of an algebra, which can be seen as a deformation of the Lie algebra~$\su(2)$.
Once an algebraic formulation is clear, this structure can be used to model a f\/inite oscillator.
The close relationship comes from the fact that for the corresponding f\/inite oscillator model the spectrum of the position operator coincides with the spectrum of the matrix~$M$.

Therefore, it is worthwhile to examine the algebraic structures behind the current matrices~$M$.
We shall do this explicitly for the three double dual Hahn cases.

For the case {\bf dual Hahn~I}, we return to the form of the matrix~$M$ given in~\eqref{MK-I} or~\eqref{M_kI}.
For any positive integer $N$, let $J_+$ denote the lower-triangular tridiagonal $(2N+1)\times(2N+1)$ matrix given below, and let $J_-$ be its transpose
\begin{gather}
\label{J+}
J_+= 2 \left( \begin{matrix}
 0 & 0 & & & \\
 M_0 & 0 & 0 & & \\
 0 & M_1 & 0 & 0 & \\
							 & 0 & M_2 & 0 & \ddots \\
 & &\ddots & \ddots & \ddots
 \end{matrix} \right),
\qquad
J_-= J_+^\dagger.					
\end{gather}
Let us also def\/ine the common diagonal matrix
\begin{gather}
J_0=\operatorname{diag}(-N, -N+1, \ldots, N),
\label{J0}
\end{gather}
and the ``parity matrix''
\begin{gather}
P=\operatorname{diag}(1,-1,1,-1,\ldots).
\label{P}
\end{gather}
Then it is easy to check that these matrices satisfy the following relations (as usual, $I$ denotes the identity matrix)
\begin{gather}
 P^2=1, \qquad PJ_0=J_0P, \qquad PJ_\pm=-J_\pm P, \nonumber\\
 [J_0, J_\pm] = \pm J_\pm, \nonumber\\
 [J_+,J_-]=2 J_0 + 2 (\gamma+\delta+1)J_0P - (2N+1)(\gamma-\delta)P + (\gamma-\delta)I.
\label{alg-I}
\end{gather}
Especially the last equation is interesting.
From the algebraic point of view, it introduces some two-parameter deformation or extension of~$\su(2)$.
When $\gamma=\delta=-1/2$, the equations coincide with the $\su(2)$ relations.
Another important case is when $\delta=-\gamma-1$, leaving a one-parameter extension of~$\su(2)$ without quadratic terms.

For the case {\bf dual Hahn~II}, the corresponding expressions of $J_+$, $J_-$, $J_0$ and $P$ are the same as above in~\eqref{J+}--\eqref{P},
but with $M_k$-values given by~\eqref{M_kII}.
As far as the algebraic relations are concerned, they are also given by~\eqref{alg-I} but with the last relation replaced by
\begin{gather*}
[J_+,J_-]=-2 J_0 + 2 (\gamma+\delta+2N+1)J_0P + (2N+1) (\gamma-\delta)P -(\gamma-\delta)I.
\end{gather*}

For the case {\bf dual Hahn~III}, the size of the matrices changes to $(2N+2)\times(2N+2)$.
For~$J_+$ and~$J_-$ one can use~\eqref{J+}, with $M_k$-values given by~\eqref{M_kIII}.
$P$ has the same expression~\eqref{P}, but for $J_0$ we need to take
\begin{gather*}
J_0=\operatorname{diag} \left(-N-\frac12, -N+\frac12, \ldots, N+\frac12\right).
\end{gather*}
With these expressions, the algebraic relations are given by~\eqref{alg-I} but with the last relation replaced by
\begin{gather}
[J_+,J_-]=2 J_0 + 2 (\gamma-\delta)J_0P - ((2N+2)(\gamma+\delta+1)+(2\gamma+1)(2\delta+1))P\nonumber\\
\hphantom{[J_+,J_-]=}{} + (\gamma-\delta)I .
\label{alg-III}
\end{gather}

The structure of these algebras is related to the structure of the so-called algebra ${\cal H}$ of the dual $-1$ Hahn polynomials,
see~\cite{Genest2013,Tsujimoto}.
It is not hard to verify that the algebra ${\cal H}$, determined by~\cite[equations~(3.4)--(3.6)]{Genest2013} or~\cite[equations~(6.2)--(6.4)]{Genest2013}, can be cast in the form~\eqref{alg-I} (or vice versa).
Indeed, starting from the form~\cite[equations~(6.2)--(6.4)]{Genest2013} coming from dual $-1$ Hahn polynomials, we can take
\begin{gather*}
J_0 = \widetilde{K_1}- \frac{\rho}{4}, \qquad J_+ = \widetilde{K_2} + \widetilde{K_3} ,\qquad J_- = \widetilde{K_2} - \widetilde{K_3},
\end{gather*}
to get the same form as~\eqref{alg-I}
\begin{gather}
 P^2=1, \qquad PJ_0=J_0P, \qquad PJ_\pm=-J_\pm P, \nonumber\\
 [J_0, J_\pm] = \pm J_\pm, \qquad
 [J_+,J_-]=2 J_0 + 2 \nu J_0P +\frac{\sigma}{2}P + \frac{\rho}{2}I,
\label{nu-rho}
\end{gather}
where $\nu$, $\sigma$, $\rho$ depend on the parameters of the dual $-1$ Hahn polynomials $\alpha$, $\beta$, $N$ through~\cite[equations~(3.4)--(3.6)]{Genest2013}.
In our case, the algebraic relations are the same, but the dependence of the ``structure constants'' in~\eqref{nu-rho} on the parameters $\gamma$, $\delta$, $N$ of the dual Hahn polynomials is dif\/ferent.

As far as we can see, the doubling of dual Hahn polynomials as classif\/ied in this paper gives a set of polynomials that is similar but in general
not the same as a set of dual~$-1$ Hahn polynomials~\cite{Tsujimoto} (except for specif\/ic values of parameters, e.g., $\delta=-\gamma-1$ does coincide with a~specif\/ic dual~$-1$ Hahn polynomial).
For general parameters, the support of the weight function is dif\/ferent, the recurrence relations (or dif\/ference relations) are dif\/ferent, and the hypergeometric series expression is dif\/ferent.

The algebraic structures obtained here (or special cases thereof) can be of interest for the construction of f\/inite oscillator
models~\cite{Atak-Suslov,Atak2001,Atak2005,JSV2011}.
Two familiar f\/inite oscillator models fall within this framework: the model discussed in~\cite{JSV2011} corresponds to~\eqref{alg-III} with $\delta=\gamma$,
and the one analysed in~\cite{JSV2011b} to~\eqref{alg-I} with $\delta=\gamma$.
Observe that there are some other interesting special values.
For example, the case~\eqref{alg-I} with $\delta=-\gamma-1$ gives rise to an interesting algebra,
and in particular also to a very simple spectrum~\eqref{324}.
We intend to study the f\/inite oscillator that is modeled by this case, and study in particular the corresponding f\/inite Fourier transform; but this will be the topic of a separate paper.

\section{Conclusion} \label{sec:conclusion}

We have classif\/ied all pairs of recurrence relations for two types of dual Hahn polynomials (i.e., dual Hahn polynomials with dif\/ferent parameters), and refer to these as dual Hahn doubles.
The analysis is quite straightforward, and the result is given in Theorem~\ref{theo2}, yielding three cases.
For each case, we have given the corresponding symmetric two-diagonal matrix $M$, its matrix of orthonormal eigenvectors $U$ and its eigenvalues in explicit form.
The same classif\/ication has been obtained for Hahn polynomials and Racah polynomials.

The orthogonality of the matrix $U$ gives rise to new sets of orthogonal polynomials.
These sets could in principle also be obtained from, for example, a~set of dual Hahn polynomials and a~certain Christof\/fel transform.
In our approach, the possible cases where such a transform gives rise to a polynomial of the same type follow naturally,
and also the explicit polynomials and their orthogonality relations arise automatically.

As an interesting secondary outcome, we obtain nice one-parameter and two-parameter extensions of the Sylvester--Kac matrix with explicit eigenvalue expressions. Such matrices can be of interest for testing numerical eigenvalue routines.

The f\/irst example of a (dual) Hahn double appeared in a f\/inite oscillator model~\cite{JSV2011}.
For this model, the Hahn polynomials (or their duals) describe the discrete position wavefunction of the oscillator,
and the two-diagonal matrix $M$ lies behind an underlying algebraic structure.
Here, we have examined the algebraic relations corresponding to the three dual Hahn cases.
It is clear that the analysis of f\/inite oscillators for some of these cases is worth pursuing.

\appendix

\section{Appendix: doubling Racah polynomials}
\label{sec:Racah}

The technique presented in Sections \ref{sec:dHahn} and \ref{sec:Hahn} is applied here for Racah polynomials.

Racah polynomials $R_n(\lambda(x);\alpha,\beta,\gamma,\delta)$ of degree $n$ ($n=0,1,\ldots,N$) in the variable $\lambda(x)=x(x+\gamma+\delta+1)$
are def\/ined by~\cite{Ismail, Koekoek,Suslov}
\begin{gather*}
R_n(\lambda(x);\alpha,\beta,\gamma,\delta)= {}_4F_3 \left( \atop{-n,n+\alpha+\beta+1,-x,x+\gamma+\delta+1}
{\alpha+1,\beta+\delta+1,\gamma+1} ; 1 \right),
\end{gather*}
where one of the denominator parameters should be $-N$:
\begin{gather}
\label{-N}
\alpha+1=-N\qquad\hbox{or}\qquad \beta+\delta+1=-N \qquad\hbox{or}\qquad \gamma+1=-N.
\end{gather}
For the (discrete) orthogonality relation (depending on the choice of which parameter relates to~$-N$) we refer to~\cite[equation~(9.2.2)]{Koekoek}
or~\cite[Section~18.25]{NIST}

Racah polynomials satisfy a recurrence relation of the form~\eqref{re} with
\begin{gather}
 y_n(x) = R_n(\lambda(x);\alpha,\beta,\gamma,\delta), \qquad \Lambda(x) =\lambda(x) = x(x+\gamma+\delta+1), \nonumber \\
 A(n) = \frac{(n+\alpha+1)(n+\alpha+\beta+1)(n+\gamma+1)(n+\beta+\delta+1)}{(2n+\alpha+\beta+1)(2n+\alpha+\beta+2)},\nonumber\\
 C(n) = \frac{n(n+\alpha+\beta-\gamma)(n+\alpha-\delta)(n+\beta)}{(2n+\alpha+\beta)(2n+\alpha+\beta+1)} . \label{Racahdata}
\end{gather}

We have applied the technique described in \eqref{yyh}--\eqref{LL}, with $y_n = R_n(\lambda(x);\alpha,\beta,\gamma,\delta) $ and
$\hat{y}_n = R_{n}(\lambda (\hat x);\hat{\alpha},\hat{\beta},\hat{\gamma},\hat{\delta})$.
The analysis is again straightforward but tedious, and the f\/inal outcome is
\begin{Theorem}\label{theoR}\allowdisplaybreaks
The only way to combine two sets of Racah polynomials such that they satisfy difference relations
of the form~\eqref{yyh},~\eqref{hhy} is one of the four cases:\\
 {\bf Racah I},
$R_n(x)\equiv R_n(\lambda(x);\alpha,\beta,\gamma,\delta)$ and $\hat R_n(x)\equiv R_n(\lambda(x);\alpha,\beta+1,\gamma+1,\delta-1)$:
\begin{gather*}
 \frac{(n+\beta+\delta+1)(n+\alpha+1)}{(2n+\alpha+\beta+2)} R_{n+1}(x) - \frac{(n-\delta+\alpha+1)(n+\beta+1)}{(2n+\alpha+\beta+2)} R_n(x)\\
 \qquad{} = \frac{(x+\delta)(x+\gamma+1)}{\gamma+1} \hat R_n(x),\\
 \frac{(n+\alpha+\beta+2)(n+\gamma+2)}{(2n+\alpha+\beta+3)} \hat R_{n+1}(x) -
 \frac{(n+1)(n-\gamma+\alpha+\beta+1)}{(2n+\alpha+\beta+3)} \hat R_n(x) \\
 \qquad{}
 = (\gamma+1) R_{n+1}(x).
\end{gather*}	
{\bf Racah II},
$R_n(x)\equiv R_n(\lambda(x);\alpha,\beta,\gamma,\delta)$ and $\hat R_n(x)\equiv R_n(\lambda(x);\alpha,\beta+1,\gamma,\delta)$:
\begin{gather*}
 \frac{(n+\gamma+1)(n+\alpha+1)}{(2n+\alpha+\beta+2)} R_{n+1}(x) - \frac{(n-\gamma+\alpha+\beta+1)(n+\beta+1)}{(2n+\alpha+\beta+2)} R_n(x)\\
 \qquad{} = \frac{(x+\beta+\delta+1)(x+\gamma-\beta)}{\beta+\delta+1}\hat R_n(x),\\
 \frac{(n+\beta+\delta+2)(n+\alpha+\beta+2)}{(2n+\alpha+\beta+3)} \hat R_{n+1}(x) -
 \frac{(n+1)(n-\delta+\alpha+1)}{(2n+\alpha+\beta+3)} \hat R_n(x) \\
 \qquad{} = (\beta+\delta+1) R_{n+1}(x).
\end{gather*}	
{\bf Racah III},
$R_n(x)\equiv R_n(\lambda(x);\alpha,\beta,\gamma,\delta)$ and $\hat R_n(x)\equiv R_n(\lambda(x-1);\alpha+1,\beta,\gamma+1,\delta+1)$:
\begin{gather*}
 \frac{1}{(2n+\alpha+\beta+2)} R_{n+1}(x) - \frac{1}{(2n+\alpha+\beta+2)} R_n(x)
 = \frac{x(x+\gamma+\delta+1)}{(\gamma+1)(\beta+\delta+1)(\alpha+1)} \hat R_{n}(x), \\
 \frac{(n+\gamma+2)(n+\beta+\delta+2)(n+\alpha+2)(n+\alpha+\beta+2)}{(2n+\alpha+\beta+3)} \hat R_{n+1}(x) \\
 \qquad \quad{}- \frac{(n+1)(n-\gamma+\alpha+\beta+1)(n-\delta+\alpha+1)(n+\beta+1)}{(2n+\alpha+\beta+3)} \hat R_{n}(x) \\
 \qquad {} = (\gamma+1)(\beta+\delta+1)(\alpha+1) R_{n+1}(x).
\end{gather*}	
{\bf Racah IV},
$R_n(x)\equiv R_n(\lambda(x);\alpha,\beta,\gamma,\delta)$ and $\hat R_n(x)\equiv R_n(\lambda(x);\alpha+1,\beta,\gamma,\delta)$:
\begin{gather*}
 \frac{ (n+\gamma+1)(n+\beta+\delta+1) }{(2n+\alpha+\beta+2)} R_{n+1}(x) - \frac{ (n-\gamma+\alpha+\beta+1)(x-\delta+\alpha+1)}{(2n+\alpha+\beta+2)} R_n(x)\\
 \qquad{} = \frac{(x+\gamma+\delta-\alpha)(x+\alpha+1)}{(\alpha+1)} \hat R_{n}( x), \\
 \frac{(n+\alpha+2)(n+\alpha+\beta+2)}{(2n+\alpha+\beta+3)} \hat R_{n+1}(x)
 - \frac{(n+1)(n+\beta+1)}{(2n+\alpha+\beta+3)} \hat R_{n}(x)
 = (\alpha+1) R_{n+1}(x).
\end{gather*}	
\end{Theorem}

Note that after interchanging $n$ and $x$, and $\alpha\leftrightarrow\gamma$ and $\beta\leftrightarrow\delta$, the relations in {\bf Racah~III} coincide with the known forward and backward shift operator relations~\cite[equations (9.2.6) and (9.2.8)]{Koekoek}.
The relations in {\bf Racah~I} were already found in~\cite[equations~(5) and (6)]{JSV2014}.

In the context of Section~\ref{sec:orthpoly} it is worth noting that the above relations also correspond to Christof\/fel-Genonimus transforms.
Taking $R_n(x)\equiv R_n(\lambda(x);\alpha,\beta,\gamma,\delta)$ in the relations~\eqref{CT}--\eqref{baC}, with data given by~\eqref{Racahdata},
the above cases {\bf Racah~I}, {\bf II}, {\bf III}, {\bf IV} correspond respectively to the choices $\nu=-\delta$, $\nu=\beta-\gamma$, $\nu=0$ and $\nu=-\alpha-1$.

For each of the four cases, one can translate the set of dif\/ference relations to a matrix identity of the form $MU=UD$. In fact, for each of the four cases, there are three subcases depending on the choice of $-N$ in~\eqref{-N}.
We shall not give all of these cases: they should be easy to construct for the reader who needs one.
Let us just give an example or two.

Consider the case {\bf Racah I} with $\alpha+1=-N$.
It is convenient to perform the shift $\delta \rightarrow \delta+1$ in the two dif\/ference relations of Theorem~\ref{theoR}.
The orthogonal matrix $U$ is of order $(2N+2)\times(2N+2)$, with elements
\begin{gather}
 U_{2n,N-x} = U_{2n,N+x+1} = \frac{(-1)^n}{\sqrt{2}} \tilde R_n(\lambda(x);\alpha,\beta,\gamma,\delta+1), \nonumber\\
 U_{2n+1,N-x} = -U_{2n+1,N+x+1} = -\frac{(-1)^n}{\sqrt{2}} \tilde R_n(\lambda(x);\alpha,\beta+1,\gamma+1,\delta), \label{R-I}
\end{gather}
where $\tilde R_n$ is the notation for a normalized Racah polynomial.
Then, one has
\begin{Proposition}
\label{propR1}
Suppose that $\gamma,\delta>-1$ and $\beta>N+\gamma$ or $\beta<-N-\delta-1$.
Let $M$ be a~tri\-diagonal $(2N+2)\times(2N+2)$-matrix of the form~\eqref{MK} with
\begin{gather*}
 M_{2k} = \sqrt{ \frac{(N-\beta-k)(\gamma+1+k)(N+\delta+1-k)(k+\beta+1)	}{(N-\beta-2k)(2k-N+1+\beta) }},\nonumber\\
 M_{2k+1} = \sqrt{ \frac{(\gamma+N-\beta-k)(k+1)(N-k)(k+\beta+\delta+2) }{(N-\beta-2k-2)(2k-N+1+\beta) }},
\end{gather*}
and $U$ the orthogonal matrix determined in~\eqref{R-I}.
Then the columns of $U$ are the eigenvectors of $M$, i.e., $M U = U D$,
where $D$ is a diagonal matrix containing the eigenvalues of~$M$
\begin{gather*}
 D= \operatorname{diag} (-\epsilon_N,\ldots,-\epsilon_1,-\epsilon_0,\epsilon_0,\epsilon_1,\ldots,\epsilon_{N}), \\
 \epsilon_{k}=\sqrt{(k+\gamma+1)(k+\delta+1)},\qquad k=0,1,\ldots,N.
\end{gather*}
\end{Proposition}

As a second example, consider the case {\bf Racah III} with $\alpha+1=-N$.
The orthogonal mat\-rix~$U$ is now of order $(2N+1)\times(2N+1)$, with elements
\begin{gather}
 U_{2n,N-x} = U_{2n,N+x} = \frac{(-1)^n}{\sqrt{2}} \tilde R_n(\lambda(x);\alpha,\beta,\gamma,\delta), \qquad n=1,\ldots,N, \nonumber \\
 U_{2n+1,N-x-1} = -U_{2n+1,N+x+1} = -\frac{(-1)^n}{\sqrt{2}} \tilde R_n(\lambda(x);\alpha+1,\beta,\gamma+1,\delta+1),\nonumber\\
\hphantom{U_{2n+1,N-x-1} = -U_{2n+1,N+x+1} =}{} n=0,\ldots,N-1, \nonumber \\
 U_{2n,N}= (-1)^n \tilde R_n(\lambda(0);\alpha,\beta,\gamma,\delta),\qquad U_{2n+1,N}=0.\label{R-III}
\end{gather}
Then, one has

\begin{Proposition}
\label{propR2}
Suppose that $\gamma,\delta>-1$ and $\beta>N+\gamma$ or $\beta<-N-\delta$.
Let $M$ be a tridiagonal $(2N+1)\times(2N+1)$-matrix of the form~\eqref{MK-I} with
\begin{gather*}
 M_{2k} = \sqrt{ \frac{(k+\gamma+1)(-N+\beta+k)(N-k)(k+\beta+\delta+1)	}{(N-\beta-2k)(N-\beta-2k-1) }},\nonumber\\
 M_{2k+1} = \sqrt{ \frac{(\gamma+N-\beta-k)(k+1)(k+\beta+1)(k-\delta-N) }{(N-\beta-2k-2)(N-\beta-2k-1) }},
\end{gather*}
and $U$ the orthogonal matrix determined in~\eqref{R-III}.
Then the columns of $U$ are the eigenvectors of $M$, i.e., $M U = U D$,
where $D$ is a diagonal matrix containing the eigenvalues of~$M$
\begin{gather*}
 D= \operatorname{diag} (-\epsilon_N,\ldots,-\epsilon_1,0,\epsilon_1,\ldots,\epsilon_{N}), \qquad
 \epsilon_{k}=\sqrt{k(k+\gamma+\delta+1)},\qquad k=1,\ldots,N.
\end{gather*}
\end{Proposition}

\subsection*{Acknowledgements}

The authors wish to thank the referees for their insightful remarks and suggestions which helped to enhance the clarity of the matter covered here.

\pdfbookmark[1]{References}{ref}
\LastPageEnding

\end{document}